\documentclass[manuscript,screen]{acmart}

\AtBeginDocument{%
  \providecommand\BibTeX{{%
    \normalfont B\kern-0.5em{\scshape i\kern-0.25em b}\kern-0.8em\TeX}}}

\setcopyright{acmcopyright}
\copyrightyear{2024}
\acmYear{2024}
\acmDOI{XXXXXXX.XXXXXXX}

\acmPrice{}
\acmISBN{}

\begin{document}

\title{Interactive $360^{\circ}$ Video Streaming Using FoV-Adaptive Coding with  Temporal Prediction}


\author{Yixiang~Mao}
\affiliation{%
 \institution{New York University}
 \streetaddress{370 Jay St}
 \city{Brooklyn}
 \state{New York}
  \country{USA}
  \postcode{11201}}
\email{yixiang.mao@nyu.edu}

\author{Liyang~Sun}
\affiliation{%
 \institution{New York University}
 \streetaddress{370 Jay St}
 \city{Brooklyn}
 \state{New York}
  \country{USA}
  \postcode{11201}}
\email{ls3817@nyu.edu}

\author{Yong~Liu}
\affiliation{%
 \institution{New York University}
 \streetaddress{370 Jay St}
 \city{Brooklyn}
 \state{New York}
  \country{USA}
  \postcode{11201}}
\email{yongliu@nyu.edu}

\author{Yao~Wang}
\affiliation{%
 \institution{New York University}
 \streetaddress{370 Jay St}
 \city{Brooklyn}
 \state{New York}
  \country{USA}
  \postcode{11201}}
\email{yw523@nyu.edu}


\begin{abstract}
For $360^{\circ}$ video streaming, FoV-adaptive coding that allocates more bits for the predicted user's field of view (FoV) is an effective way to maximize the rendered video quality under the limited bandwidth. We develop a low-latency FoV-adaptive coding and streaming system for interactive applications that is robust to bandwidth variations and FoV prediction errors. 
To minimize the end-to-end delay and yet maximize the coding efficiency, we propose a frame-level FoV-adaptive inter-coding structure. 
In each frame, regions that are in or near the predicted FoV are coded using temporal and spatial prediction, while a small rotating region is coded with spatial prediction only. This rotating intra region periodically refreshes the entire frame, thereby providing robustness to both FoV prediction errors and frame losses due to transmission errors.  
The system adapts the sizes and rates of different regions for each video segment to maximize the rendered video quality under the predicted bandwidth constraint. 
Integrating such frame-level FoV adaptation with temporal prediction  is challenging due to the temporal variations of the FoV.  We propose novel ways for modeling the influence of FoV dynamics on the quality-rate performance of temporal predictive coding.
We further develop LSTM-based machine learning models to  predict the user's FoV and network bandwidth.
The proposed system is compared with three benchmark systems, using real-world network bandwidth traces and FoV traces, and is shown to significantly improve the rendered video quality, while achieving very low end-to-end delay and low frame-freeze probability. 
\end{abstract}




%%
%% Keywords. The author(s) should pick words that accurately describe the work being presented. Separate the keywords with commas.

\maketitle

%------------------------------------------------------------------------- 
\section{Introduction}
\label{sec:introduction}

Effectively coding and streaming $360^{\circ}$ video is critically important  for Virtual Reality (VR) and Augmented Reality (AR) applications. However, to achieve similar viewing quality, the required network bandwidth for sending the omni-directional video is much higher than that required for the traditional planar video. For example, to provide a similar premium quality as 8K ($7680\times4320$ pixels) planar video, the $360^{\circ}$ video need to have 24K ($23040\times11520$ pixels) resolution . Previous study \cite{huawei_report} showed that transmitting such high-resolution video at 120 frames per second (fps) easily consumes Gigabits-per-second bandwidth.
{\it FoV-adaptive streaming} is an effective way to reduce such high bandwidth requirement \cite{8931627}\cite{9238515}.
In a $360^\circ$ video session, a viewer only watches the content within a limited Field-of-View (FoV) at any time, which is a small portion of the $360^\circ$ scope. A viewer's FoV in each frame can be predicted using various methods. FoV-adaptive streaming leveraging tile-based coding refers to the strategy that encodes and transmits the tiles inside the predicted FoV at premium quality, while discarding or encoding and transmitting the remaining tiles at significantly lower quality. Such strategy has been widely developed and evaluated in $360^\circ$ video-on-demand systems \cite{8931628,fov_adapt_2,fov_adapt_3,1,tile_based_3,qian2016optimizing,hou2020predictive} and $360^\circ$ live streaming systems\cite{8999740, live_1,live_2,live_3, sun2020flocking, 8016378}.
However, this approach has not been sufficiently investigated for interactive applications, such as VR cloud gaming, VR video conferencing, and AR remote collaboration, etc.  \cite{interactive_gamming, vr_conferencing, lee2015outatime}.

A major challenge in interactive applications is that $360^{\circ}$ video must be coded and transmitted in real-time with extremely low latency (e.g. $\leq$100ms).
% Motion-compensated temporal prediction has been widely used in planar video coding to maximize the video quality while significantly reduce the bandwidth requirement. 
%The commonly used group of pictures (GOP) structure is not appropriate for interactive applications. Its periodic intra frames require much more bits than inter frames, leading to periodic large bit rate spikes, which in turn dramatically increase the overall system latency.
\textbf{To achieve low latency, $360^{\circ}$  video should be coded and delivered at the frame level, and each frame should consume similar rates.}  To accomplish this, previous work on  interactive $360^\circ$ video  \cite{interactive_gamming,pitkanen2019remote} intra-code all frames without using motion-compensated temporal prediction. A big drawback of using such intra-only coding mode is that it significantly reduces the coding efficiency, which translates to significantly lower  video quality under the same bandwidth.
On the other hand, integrating temporal predictive coding with frame-level FoV adaptation  faces the following challenges:

\begin{enumerate}
\item The periodic intra-frame structure is not suitable for the interactive applications, because it leads to periodic rate spikes and consequently increased delay. On the other hand, it is important to periodically update the entire $360^\circ$ scope to limit error propagation after a frame loss due to transmission errors, and to mitigate the quality degradation in un-coded regions, which in turn affects temporal prediction accuracy for future frames.
\item The coded regions in each frame depend on the predicted FoV. User FoV in successive frames are often not aligned. Such misalignment causes prolonged time lapse for temporal prediction for some tiles and leads to reduced coding efficiency. How to properly consider such reduced coding efficiency is important for accurate rate control and essential for minimizing the latency, while optimizing for the rendering quality.  
\end{enumerate}

% Without using temporal prediction (i.e. using only intra-coding for all the frames, as in some prior work \cite{interactive_gamming}\cite{pitkanen2019remote}), it is much easier to perform FoV adaptive coding and rate control, but it would lead to significantly lower quality under the same network throughput.

\begin{figure}
\centerline{\includegraphics[scale=0.2]{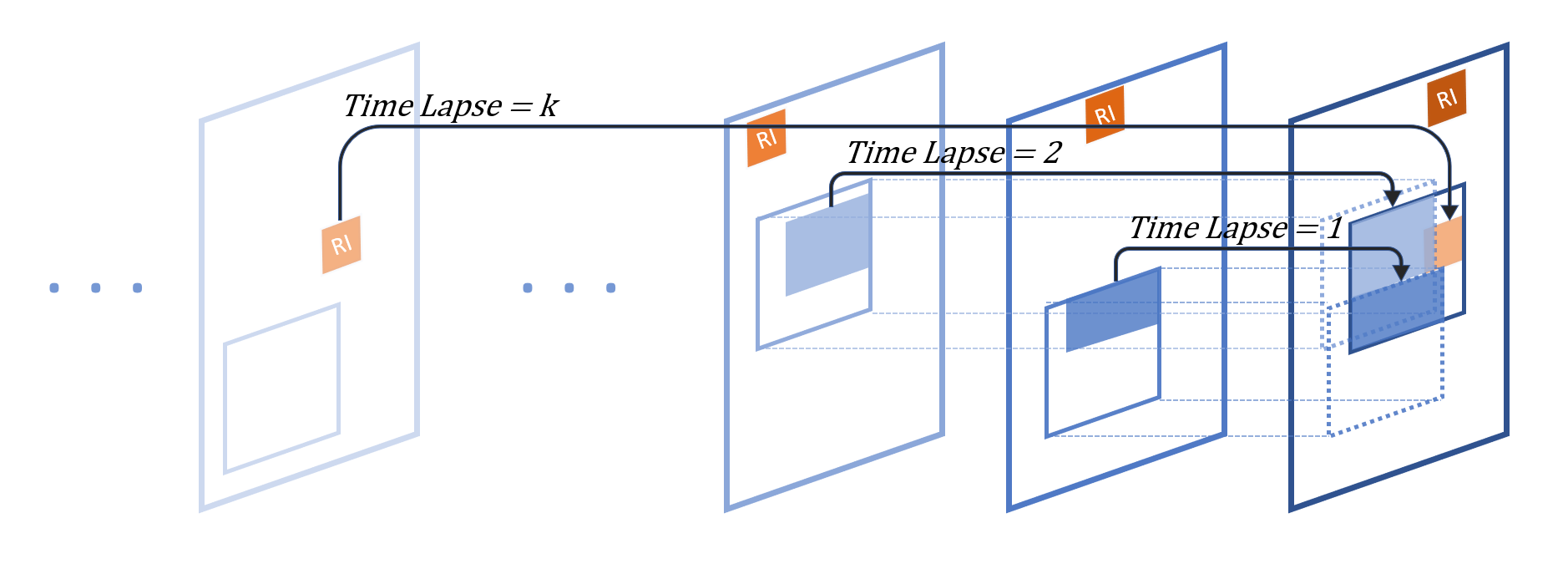}}
% \vspace{-3mm}
\caption{Variable time lapses between the coded tiles inside the PF and PF+ regions. The frame on the right is the current frame, its previous frames are on its left. The square region covered by a solid-line border in each frame indicates the coded region in that frame. Different tiles in the coded region in the current frame have different time lapses to the latest frame when the corresponding tiles were coded. }
\label{fig:codeing_dependency}
\vspace{-5mm}
\end{figure}

This paper presents a novel low-latency FoV-adaptive coding and streaming solution for interactive $360^{\circ}$ video.
% We assume the sender and the receiver are connected by a network path with dynamically varying throughput without short-latency guarantee. 
% The sender is either the video source, or a proxy server relaying the source video. 
% The receiver is either the end user device that directly renders the video, or a local edge server that renders the video and transmit to the end user \cite{Hou2017}.
We adopt motion-compensated temporal predictive coding to maximize the quality-rate efficiency, and address the challenges brought by temporal prediction in frame-wise FoV-adaptive coding. 
To reduce the system latency, we propose using a rotating-intra region in each frame to replace the periodic intra-frame. We also explicitly model the impact of the misalignment of coded regions on the quality-rate performance of temporal predictive coding, to enable accurate rate allocation among different regions. 
%A novel LSTM fusion model is also proposed to accurately predict the network bandwidth. 
The salient features of our proposed solution include:

\begin{enumerate}          
\item To achieve low latency, the sender codes and transmits video at the granularity of frames, instead of segments (e.g. groups of pictures) commonly used for video-on-demand and live streaming.
\item For each new frame, the sender predicts the FoV of the receiver, and codes only a region covering the predicted FoV (denoted PF) plus a surrounding border (denoted PF+), with the border size adapted to the  anticipated FoV prediction errors. Both regions will be coded using temporal prediction but at different rates. \footnote{When a remote site has multiple participants, we can take the union of the FoVs of all the participants as the ground truth ``FoV'' of this site, and predict the future FoV union.}
%{Typically, all users tend to focus on the same region of interest (e.g. the speaker in a conference), so the FoV union may be only slightly larger than the single user's FoV. Occasionally, if the users' FoVs diverge significantly, we have to code a larger region, making FoV adaptive coding less efficient. }
\item We code a small rotating region using intra-coding for each frame using spatial prediction only, enabling gradual refreshment of the entire $360^\circ$ scope after a certain period. See Fig.~\ref{fig:frame_structure}. Such rotating-I (RI) regions reduce the frame size burstiness and hence reduces delay, while providing robustness against FoV prediction errors and frame losses.
\item While modeling the quality-rate (Q-R) relations of coded regions, we take into account the spatially and temporally varying time lapses in temporal prediction due to FoV dynamics (see Fig.~\ref{fig:codeing_dependency}). 
%This is accomplished  We model the rate-increase factor as a function of the time lapse  and use the time lapse distribution to compute the expected rate increase. 
We further model the decay of the rendering quality of non-coded regions  as a function of the time lapse since these regions were last  coded.
\item We periodically adapt the sizes and the target normalized bit rates of different coding regions at the segment level,  based on the predicted network bandwidth and FoV prediction accuracy, guided by the developed Q-R models.
\item We develop LSTM-based deep learning models for frame-level FoV prediction and segment-level bandwidth prediction, respectively.
%\item We propose a novel LSTM fusion model for bandwidth prediction that adaptively combines predictions by models trained with different temporal history lengths, leading to more accurate prediction in dynamically changing network environments than using a fixed history length. 
The FoV and bandwidth prediction modules in the streaming system can be replaced by more powerful prediction algorithms in the future.
\item To avoid self-congestion, we design push-based frame delivery scheme with short sender and receiver buffers. We further adjust the frame-level bit budget in real-time and control sender buffer overflow, to maximize the frame delivery rate before the display deadline.
\end{enumerate}

The proposed solution is simulated and compared with three benchmark systems, using the FoV traces and network bandwidth traces collected in real systems. 
The simulation results show that the proposed system achieves significantly higher rendered quality than all the benchmark systems, especially the benchmarks only using intra-coding. Compared to the inter-coding benchmark (using periodic I-frames), the proposed system has much lower delay and less freeze. We further simulated a  simplified version of the proposed system and the results show that the benefit from region size adaptation is marginal. Hence, the complexity of the proposed system can be reduced in practice by fixing the region sizes. 

Sec.~\ref{sec:related_work} summarizes the related works on FoV-adaptive $360^\circ$ video streaming.
Sec.~\ref{sec:tile} introduces the proposed tile-based video coding scheme including the frame partition scheme and the optimization of the  tile size. 
Sec.~\ref{sec:video_coding_section} considers how to model the rate-distortion functions of the proposed coding scheme in the presence of temporal variation of the FoV on and furthermore how to optimize the region size and rate given target total bit rates. 
Sec.~\ref{sec:fov_bd_section} describes the deep learning models used for bandwidth prediction and FoV prediction.
Sec.~\ref{sec:streaming_section} presents the push-based FoV-adaptive streaming system, including the adaptation of region size and rate allocation at the segment level and the bit budget adjustment at the frame level.
Sec.~\ref{sec:simulation_exp} explains the setup of the trace-driven simulations, describes the evaluation metrics, and compares system performance with benchmarks using intra- or inter-coding. The last section summarizes the contributions and takeaways of our work.

Preliminary results of this work were presented in \cite{mao2020low}, which used simpler approaches for FoV and bandwidth prediction. 
The current system using LSTM-based models enjoys higher FoV and bandwidth prediction accuracy, which in turn leads to reduced latency and freeze and improved rendering quality, the update results are shown in Sec.~\ref{sec:simulation_exp}. 
Compared to the previous paper \cite{mao2020low}, Sec.~\ref{sec:related_work}, Sec.~\ref{sec:optimal_tile_size}, and Sec.~\ref{sec:fov_bd_section} are newly added. This paper also includes more details about video coding experiments in Sec.~\ref{sec:tile} and Sec.~\ref{sec:video_coding_section}.

%------------------------------------------------------------------------- 
\section{Related Works}
\label{sec:related_work}

% \subsection{FOV-adaptive $360^{\circ}$ video streaming}
% \label{sec:related_fov_adap}

Given the nature that a viewer can only watch a small portion (Field of View, or FoV) of the entire $360^{\circ}$ scope at any time, 
most of recent $360^{\circ}$ video streaming systems adapt the coding and transmitting areas based on the predicted viewer's FoV \cite{7501810,1,pred1,ban2018cub360,pred2,live_1,sun2022live,sun2020flocking,interactive_gamming,pitkanen2019remote}.
% , for both video-on-demand (VoD) [][][][][][] and live streaming applications [][][][][][]. 
These streaming systems usually encode and transmit the video content inside the predicted FoV at premium quality, while discarding or encoding and transmitting the content outside the predicted FoV at significantly lower quality. In the following, we  review prior work in $360^{\circ}$ video streaming for three different application scenarios separately.

For video-on-demand (VoD) applications, which can afford a relatively long latency (e.g. over 20 sec.), a video is typically divided into segments (each  1 sec. or longer) and the FoV adaptation is done at the segment level. That is, the FoV distribution for a entire future segment is predicted, and the same rate allocation across the  $360^{\circ}$ scope is used for the entire segment. 
In \cite{7501810}, Auto Regressive Moving Average (ARMA) is used to predict the future viewing position. The $360^{\circ}$ scope is divided into non-overlapping tiles, and different quality levels are assigned for different tiles.
In\cite{1},  a 2-tier design is proposed. The base tier includes the entire $360^{\circ}$ scope, whereas the enhancement tier covers only an enlarged portion of the predicted FoV. The base layer is always delivered so as to provide acceptable quality even in the presence of FoV prediction error. Furthermore the base layer is prefetched to guarantee its delivery. Truncated Linear Prediction (TLP) model is used to predict the viewer's future viewing trajectory using the past trajectory.
In \cite{pred1}, linear regression is proposed to predict not only the viewers' viewing trajectory, but also the deviation of the predicted trajectory. Such deviation helps the system to decide how large a margin around the predicted FoV region is needed in case the prediction is not accurate.
\cite{ban2018cub360} combines the current viewer's FoV with cross-users behaviors to predict the current viewer's future FoV, and then tiles the frame for rate allocation. 
In \cite{fov_pred}, the authors propose to use LSTM-based models that uses the viewer's past trajectory as well as other viewers' future viewing trajectory, and is successful in predicting a viewer's viewing trajectory over a long time horizon, which is necessary for VoD systems.  
Image saliency maps and motion maps are also utilized to predict future FoV in \cite{pred2}. 

Live $360^{\circ}$ streaming applications have a shorter latency requirement (e.g. up to 10 sec.) and  have to serve multiple clients, and hence is more challenging than VoD.
Several prior works design the FoV adaptation mechanisms specifically for live streaming.  Instead of predicting the center of the FoV, \cite{live_1} proposes to predict the likelihood of different areas being viewed, and a scalable real-time coding and multicasting framework is proposed.
The authors of \cite{sun2022live} and \cite{sun2020flocking} propose to use the viewing traces of viewers with a shorter latency to help the FoV prediction of the viewers with longer latency, based on the fact that viewers usually focus on several attractive spots in the $360^{\circ}$. They propose a light-weight LSTM-based model that works with a variable number of users and improves the prediction accuracy significantly for the viewers having longer latency.

FoV-adaptive streaming of $360^{\circ}$ videos for interactive applications have unique challenges. Since the interactive applications require a very short latency (one the order of 100 ms.), the $360^{\circ}$ video has to be coded with the FoV adaptation at the frame level. Because the viewer's FoV in successive frames are often not aligned, the location of coded tiles vary from frame to frame. This makes utilizing motion-compensated temporal prediction for coding the video tiles extremely challenging.
Prior work on  FoV-adaptive streaming for interactive applications are quite limited. To avoid the difficulty caused on the temporal variation of the FoV, \cite{interactive_gamming}\cite{pitkanen2019remote} intra-code all vertical slices centered at the predicted FoV center in each frame without using temporal prediction, which leads to significantly reduced coding efficiency and consequently lower video quality under the same bandwidth. Also, by dividing the frame in Equirectangular projection (ERP) using vertical slices, their system has to encode and transmit all slices in order to cover the FoV when the viewer watches the pole directions.

To avoid the coding efficiency loss, we apply motion-compensated temporal predictive coding, while taking into account of the misalignment of the FoV regions in successive frames. We adopt the tile-based frame structure in our interactive streaming system, and allocate rates for tiles based on their likelihood to be viewed. Instead of using periodic intra-frames (Benchmark 3), which can lead to rate spikes when coding the I-frame, we use rotating intra-slice to  maintain low latency. Our streaming system significantly improves the coding efficiency of benchmark systems that use only intra-coding \cite{interactive_gamming,pitkanen2019remote} while preserving similar  end-to-end latency and freeze. 
% Preliminary results of this work were presented in \cite{mao2020low}, which used simpler approaches for FoV and bandwidth prediction. The current system using LSTM-based models enjoys higher FoV and bandwidth prediction accuracy, which in turn leads to  reduced latency and freeze and improved rendering quality.  This paper  also includes more details about video coding experiments.

% FoV prediction for 360◦ video streaming applications:
% video-on-demand (VoD): [1], [7], [8], [9], [10], [11], [12], [13], [14]
% live: [15], [16], [17]
% interactive: [18], [19], [20]. 

% \subsection{Interactive video streaming}
% \label{sec:related_interactive}

%------------------------------------------------------------------------- 
\section{Proposed FoV-Adaptive Coding Scheme}
\label{sec:tile}

\subsection{Tile-based Frame Partitioning and Rate Adaptation Based on Predicted FoV}

\begin{figure}[t]
\centerline{\includegraphics[scale=0.15]{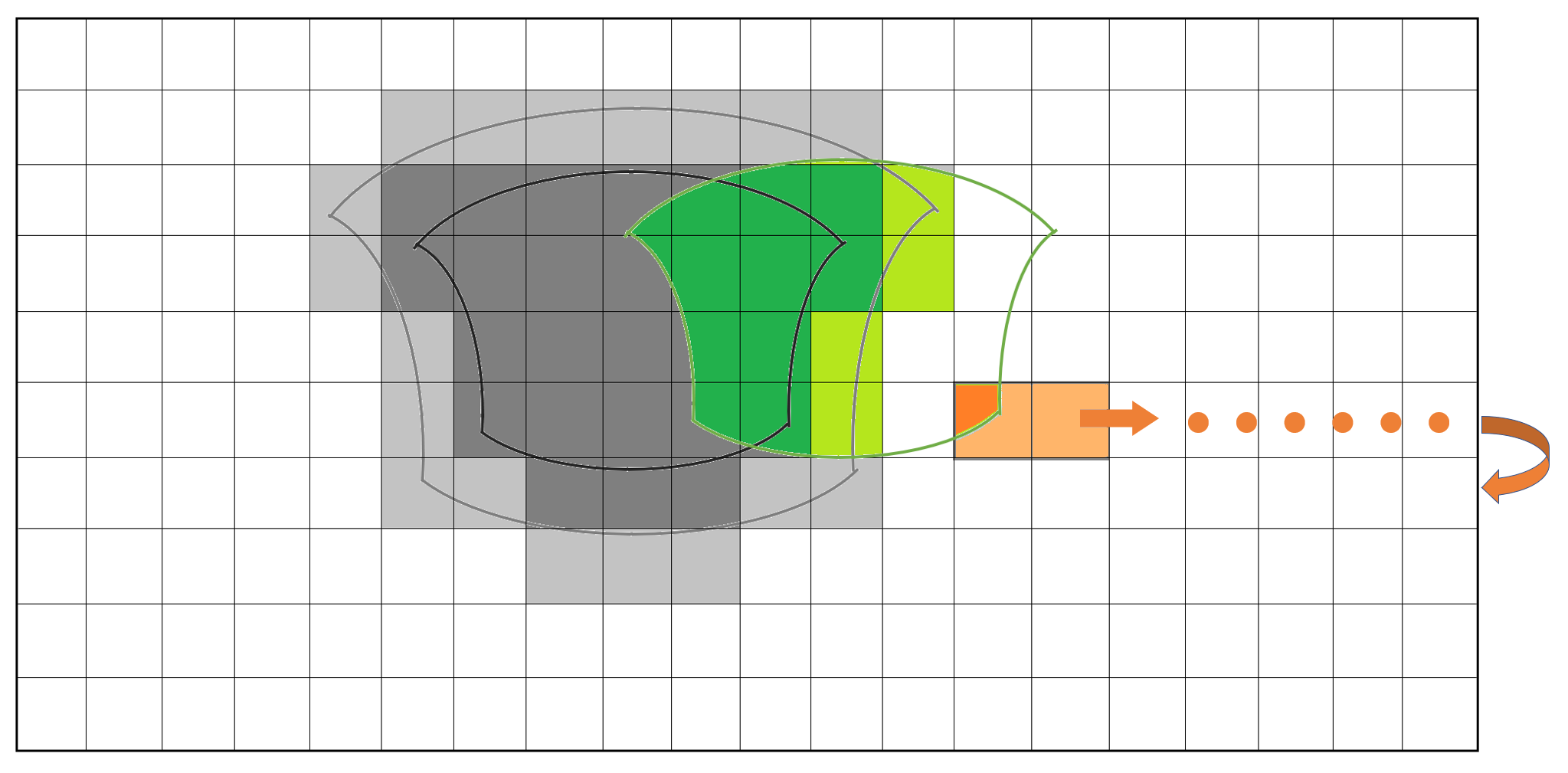}}
\vspace{-1mm}
\caption{The tiled ERP frame and different coding regions. Dark grey: tiles to cover the PF region, coded at the rate $R_e$. Light grey and orange: tiles to cover PF+ and RI, coded at the rate $R_b$. Green: user's actual FoV, which may intersect with PF, PF+, RI, and un-coded tiles. 
%The hit rate for the PF is the percentage of pixels in the rendered FoV that are in the PF tiles. One can similarly define the hit rates of other regions.
} 
\label{fig:frame_structure}
\vspace{-6mm}
\end{figure}

We propose a novel tile-based FoV-adaptive coding structure to replace the conventional group of pictures (GOP) structure, which uses periodic intra frames, for coding the ERP $360^{\circ}$ video. In the proposed structure, only the first frame of a video stream was encoded entirely using spatial prediction (i.e. intra-coding) only. For each following frame, we first predict the client's FoV at that time. As illustrated in Fig.~\ref{fig:frame_structure}, we then code the predicted FoV region (called ``PF") and a small border around it (called ``PF+") on the ERP using temporal prediction (i.e. inter-coding) based on the previously decoded reference frame. The border region is coded in case the PF is slightly off from the client's actual FoV. The normalized bit rate (in terms of bits/degree$^2$) allocated for the PF+ is lower than that for the PF. In addition to the PF and PF+, we also code and send a rotating-I region (called ``RI") using spatial prediction only, to ensure all pixels on the ERP will be  refreshed by intra coding at a certain frequency. 
% prevent any regions on the ERP from possibly never getting updated. 
For each successive frame, the intra-coded RI region rolls to a new location on the EPR (from top to bottom and left to right).
% The RI region refreshes all pixels on the ERP at a certain rolls in successive frames from top to bottom and left to right in the ERP. The RI is introduced to ensure that all pixels in the ERP will be refreshed using intra-coding after a certain period.  
For instance, if the size of the RI region is 1/36 of the ERP frame, the RI will refresh the entire $360^{\circ}$ scope  every 36 frames.
This periodic refreshment makes the system robust to both FoV prediction errors and frame losses due to packet losses. Since the RI region has a low chance to be viewed, it is allocated with a rate lower than PF and PF+ regions.
Note that the first frame (entire $360^{\circ}$ scope) needs to be intra-coded at a lower rate (high quantization level) to reduce the bits of the initial frame, and hence reduce the initial buffering time.

We use  tile-based coding to facilitate  adaptation of  coding methods and rates for different regions \cite{7501810,9220800}. In tile-based  $360^{\circ}$ video  coding, the entire $360^{\circ}$ scope in the ERP is partitioned into non-overlapping tiles \cite{7501810,9220800}.  Each  tile is
coded and transmitted independently.
All tiles covering the PF region are inter-coded at the normalized rate $R_e$, and all the remaining tiles covering the PF+ region are inter-coded at the normalized rate $R_b$. Note that the shape of a PF or PF+ region on the ERP depends on its latitude, as shown in Fig.~\ref{fig:frame_structure}. Therefore, the number of tiles needed to cover the PF or PF+ region may differ in each frame. 
On the other hand, the RI is a rectangular region on the ERP consisting of a fixed number of tiles, irrespective of FoV. 
To simplify the rate allocation, we code the RI and PF+ region using the same normalized rate $R_b$. Since the RI uses intra-coding and the PF+ uses inter-coding, the quality of RI is lower than the quality of PF+, even though they share the same average allocated rate $R_b$. Because the RI region rotates on the entire ERP, it may have overlap with the PF or PF+ region in some frames. Those tiles in the RI that fall within the PF or PF+ region are treated as RI and coded using the intra-mode, in order to eliminate the decoding error propagation caused by potential frame losses.
% after the intra refreshment period.
% In the decoder, which is also part of the encoder to derive the reference frame for temporal prediction, only coded tiles (those in the PF, PF+, and RI region) will be updated based on the received bits for this frame. 
The decoder only decodes and updates the tiles in the PF, PF+, and RI regions. In the un-coded regions, content from the latest decoded frames will be replicated. Other more sophisticated error concealment methods can be  incorporated in the future to enhance the quality of these un-coded regions.

The proposed system adapts the region sizes and rates for each video segment (1 sec. long in our experiments) based on the predicted network throughput. The adaptation is optimized utilizing the expected quality-rate (Q-R) functions in PF, PF+ and RI regions, which will be discussed in Sec.~\ref{sec:video_coding_section}.

\subsection{ Optimization of Tile Size}

\label{sec:optimal_tile_size}

\begin{figure}[h]
  \centering
  \includegraphics[width=\linewidth]{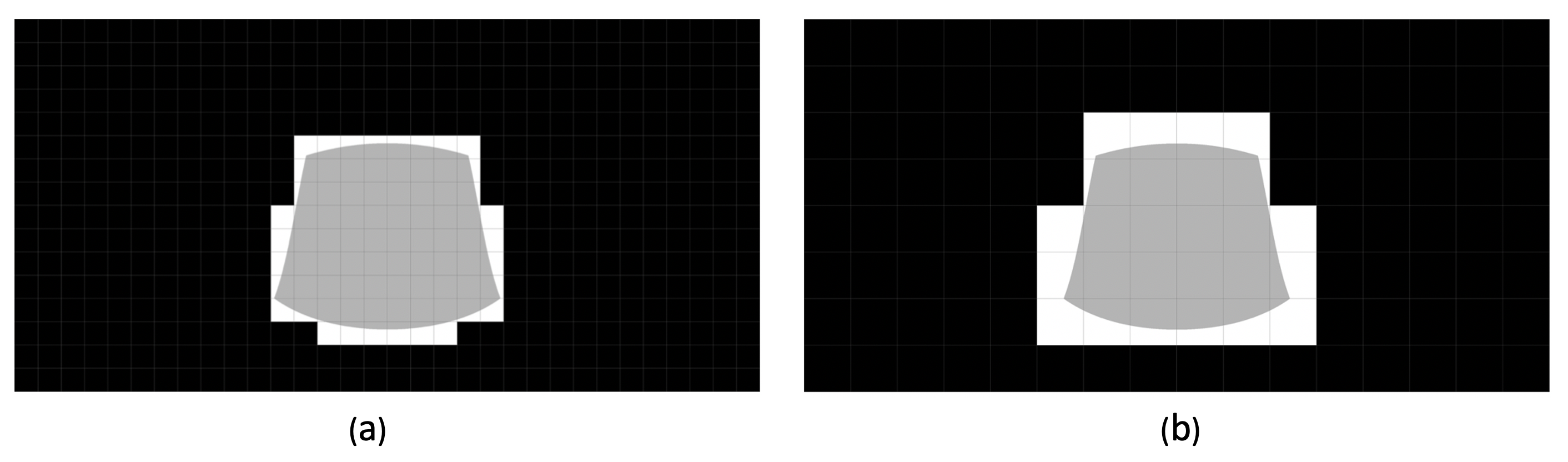}
  \caption{Tiles needed to cover the same FoV. The grey area indicate FoV, which in this example covers 90 degree. White and grey areas indicate all the titles that are needed to cover the FoV.}
  \label{fig:small_tile}
%   \Description{A woman and a girl in white dresses sit in an open car.}
\end{figure}

In tile-based $360^{\circ}$ video coding,  the tile size affects both the video coding efficiency and transmission flexibility. Since each tile is coded independently and can only access the spacial information from the tile itself, larger tile sizes generally provide higher coding efficiency.
However, a larger tile size also leads to more unused areas outside the FoV and the bits for coding those unused areas are wasted, as shown in Fig. ~\ref{fig:small_tile}. The optimal tile size should minimize the total bit consumption of all tiles needed to cover the FoV averaged over all possible viewports for a fixed quality.

\begin{figure}
\centerline{\includegraphics[width=0.5\columnwidth]{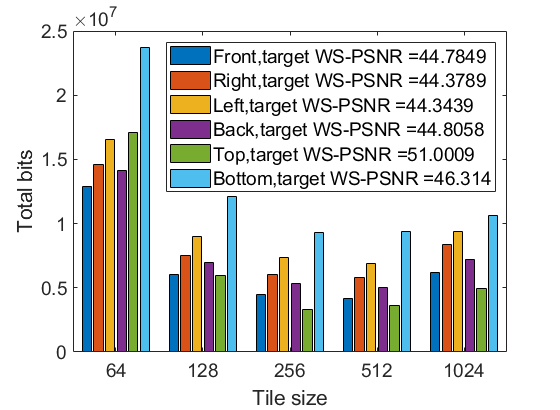}}
\caption{Total bit consumption under the same QP inside the FoV for 5 different tile sizes for 8K video. The horizontal axis indicates the number of pixels in each side of the square tile. 
All the tiles are coded in the inter-coding mode except the first frame. A constant QP=30 was used. The resulting WS-PSNR in the FoV region are reported in the figure legends. Different color bars are results for FoVs in different directions. The results are for sequence ``Trolley", similar trends are observed for ``Chairlift".}
\label{fig:optimal_tile}
\end{figure}

Early work only explored this trade-off for 1080p and 4K videos \cite{3}. We conduct a similar experiment for the 8K resolution JVET $360^{\circ}$ test videos (the detail of the video coding set-up and the test sequences is described in Sec.~\ref{sec:simulation_exp}). 
Fig.~\ref{fig:optimal_tile} illustrates the total bit rates needed when the tiles are coded in the inter-coding mode at a constant quantization parameter (QP) to cover a $90^{\circ}\times90^{\circ}$ FoV for 5 different tile sizes. Generally, the number of tiles needed differs depending on the viewport direction. 
We found for the 4 FoVs centered on the equator (front, right, left, and back), they achieve minimal bit rate consumption when using tiles of size $512\times512$ pixels, slightly better than using a tile size of $256\times256$ pixels. For the 2 FoVs facing the top or bottom directions, there are relatively more wasted pixels in the boundary tiles. In this case, $256\times256$ tile size is slightly better than $512\times512$. 
% To simplify the system setting, we pick one tile size for the entire ERP instead of coding and transmitting different regions using 2 different tile sizes. 
Given that a smaller tile size offers more granularity in varying the sizes of RI and PF+, we choose $256\times256$ pixels as the tile size for the 8K video sequences. Note that all coding experiments and streaming simulations in this paper are conducted using this tile size.

%------------------------------------------------------------------------- 
% \section{Frame-level FOV-adaptive $360^{\circ}$ video coding using spatial and temporal prediction}
\section{ Rate Distortion modeling and Adaptation of Region Size and Rate}
\label{sec:video_coding_section}

\label{sec:qr-adjusted}
To facilitate the adaptation of the region sizes and rates for each video segment (1 sec. long in our experiments) given the target total bit rate, we need to have accurate models of the  quality-rate (Q-R) functions in PF, PF+ and RI regions. The main challenge to model those Q-R relations is that the FoV location and consequently the coded regions vary temporally on the ERP. In this section, we first introduce the objective quality metric for the $360^{\circ}$ video. 
Then,  we  model the ``ideal'' Q-R functions assuming  the user's FoV does not change during the entire duration of the video sequence. 
Next, we  consider how to adjust the resulting Q-R functions to take into account of the dynamics of the FoV.
%Since the FoV and hence the coded regions may change in each frame, some tiles may be predicted from un-coded regions in the previous frame in real applications. 
%The quality of these un-coded regions depends on the time lapse since they were last coded. We introduce the rate-increase factor to account for such time lapse for temporal prediction. 
%Furthermore, we use the quality-decay factor to model the degraded quality of the non-coded tiles for the present frame.
Finally, we present a solution for optimization of the region size and rate using our rate-distortion models.

\subsection{Objective quality metric}
Weighted-to-spherically-uniform peak-signal-to-noise ratio (WS-PSNR) is an objective quality metric to evaluate the $360^{\circ}$ video recommended by JVET\cite{wspsnr}. WS-PSNR is defined as: 
 %\vspace{-2mm}
 \begin{equation}
 \label{eq:wspsnr}
    \mathrm{\mbox{WS-PSNR}} = 10\log_{10}\frac{\mathrm{MAX_I}^2}{\mathrm{\mbox{WS-MSE}}},
 \end{equation}
with  %\vspace{-3mm}
 \begin{equation}
 \label{eq:ws_wse}
    % \mathrm{\mbox{WS-MSE}} = \frac{1}{N_p}\sum_{i,j}[I(i,j)-K(i,j)]^2\cos ((\frac{i}{m}-\frac{1}{2})\pi),
    \mathrm{\mbox{WS-MSE}} = \frac{1}{\sum_{i,j}w(i,j)}\sum_{i,j}[I(i,j)-K(i,j)]^2w(i,j),
    %\vspace{-1mm}
 \end{equation}
where $i,j$ is the coordinate of a pixel on the ERP frame, $K(i,j)$ and $I(i,j)$ are the color intensity of the pixel $(i,j)$ on the raw and the encoded ERP frames, respectively. The weight $w(i,j)=\cos ((\frac{i}{m}-\frac{1}{2})\pi)$ is a factor to model the geometric distortion due to ERP projection along the pitch axis.
To calculate the quality of pixels inside an actual rendered FoV, we calculate WS-MSE of pixels inside the projected FoV on the ERP and then derive the WS-PSNR of the FoV. This metric is used to measure the rendered quality. The same method is applied to calculate the quality of the PF, PF+, RI, or the remaining regions.

\subsection{``Ideal'' Quality-Rate Models For Different Coded Regions}
\label{sec:qr}

\subsubsection{Quality-Rate Function for the Predicted FoV region}
\label{sec:qr_pf}
Because the shapes and consequently the Q-R relations of the PF and PF+ regions depend on the FoV location, we first empirically determine these Q-R relations separately for six different FoV locations: front, left, right, back, top and bottom. To model the ``ideal'' Q-R functions, we fix the FoV locations through the entire sequences, so that all tiles in the regions are continuously updated in each frame. We select two JVET $360^{\circ}$ test sequences to represent different video content: one stable video shot by a fixed camera and another dynamic video captured by a moving camera. We conduct the coding experiments to derive the Q-R functions for these videos.
The reference HEVC Test Model (HM) software~\cite{hevc} is used under JVET common test condition (CTC). Each tile in the sampled PF and PF+ regions is encoded independently using low delay P (LDP) configuration with IntraPeriod = -1, meaning only the first frame is I frame and all the  following frames are P frames. Each tile from the RI region is encoded using intra-only configuration with IntraPeriod = 1, meaning every frame is coded as I frame.
We determine the rate and the corresponding quality (WS-PSNR) under four different quantization parameters (QP): ${27, 32, 37, 42}$.
% We calculate the total rate of tiles covering the PF and PF+ regions, along with the WS-PSNR of that region, to fit the empirical Q-R curves for selected six FoV locations. 
Then, to generate the Q-R curve for a FoV region, we determine the tiles needed to cover this FoV, and calculate the WS-PSNR for all the pixels inside the FoV and the total bits of all tiles needed to cover the FoV, for each QP. 
Figures~\ref{fig:qr}(a) and \ref{fig:qr}(e) each shows six curves for the six sample FoV orientations. Here we assume the FoV size is $90^{\circ}\times90^{\circ}$. The normalized bit rate (bits/degree$^2$) is determined by dividing the total bits by $90^{\circ}\times90^{\circ}$. We further determine the average ``ideal" Q-R curve for PF, by averaging the Q-R functions for the six FoVs, based on their probabilities. 
From the statistics of the viewers' FoV behavior \cite{6}, more than 90\% FoV centers are located in the range of equator$\pm 45^{\circ}$. Therefore, we assume that the probabilities for watching the front, left, right, back, top, and bottom FoV are $0.2, 0.2, 0.2, 0.2, 0.1, 0.1$, respectively. The average Q-R curves are also shown in Fig.~\ref{fig:qr}(a) and \ref{fig:qr}(e). Finally, we approximate the weighted average Q-R curve by a logarithmic model:
 \begin{equation}
    Q_{\rm PF}(R) = a_{\rm PF} + b_{\rm PF} \log R.
 \end{equation}
As indicated in the figure legends, the model parameters $a_e$ and $b_e$ are generally content-dependent. Because "Chairlift" video contains more dynamic  motion, temporal prediction in inter-coding is more challenging, leading to lower average Q-R curves than those for "Trolly".

\subsubsection{Quality-Rate Functions for the PF+ Region}
\label{sec:qr_pf+}

As shown in Fig.~\ref{fig:frame_structure}, the PF+ region covers a border outside the PF region, and the number of tiles needed to cover the PF+ region depends on the width of the border (in degree). In our experiments, we set the border width to $ {10^{\circ},20^{\circ},30^{\circ},40^{\circ},50^{\circ}}$ ($10^{\circ}$ means that the extended degree in each direction is $5^{\circ}$). For each of the six FoV orientations, we determine the WS-PSNR among pixels falling in the border region, and count the total number of bits used by the tiles needed to cover the PF+ region, for each target PF+ size in degree. The normalized rate is determined by dividing the total rate by the border size in square degree. For example, with FoV size of $90^{\circ}\times90^{\circ}$, and border size of $10^{\circ}$, the border area is approximated by $100^{\circ}\times100^{\circ} - 90^{\circ}\times90^{\circ}$. The Q-R plots of the six FoV orientations and the weighted average Q-R curve for the border width of $10^{\circ}$ are shown in Figs.~\ref{fig:qr}(b) and \ref{fig:qr}(f). Figures~\ref{fig:qr}(c) and \ref{fig:qr}(g) show the average Q-R curves for different border widths. 
We find that the Q-R curves for different PF+ regions can also be approximated well by the logarithmic model:
 \begin{equation}
    Q_{\rm PF+}(R) = a_{\rm PF+} + b_{\rm PF+} \log R,
 \end{equation}
where the model parameters depend on the PF+ border size. Figure \ref{fig:qr}(c) and \ref{fig:qr}(g) show the average Q-R curves for different border widths. 
Note that the coding efficiency is higher for a wider border due to the fact that lower percentage of pixels in the coded PF+ tiles are wasted in such a case.

\subsubsection{Quality-Rate Functions of the RI Region}
\begin{figure*}[t]
    \centering
    \includegraphics[scale=0.17]{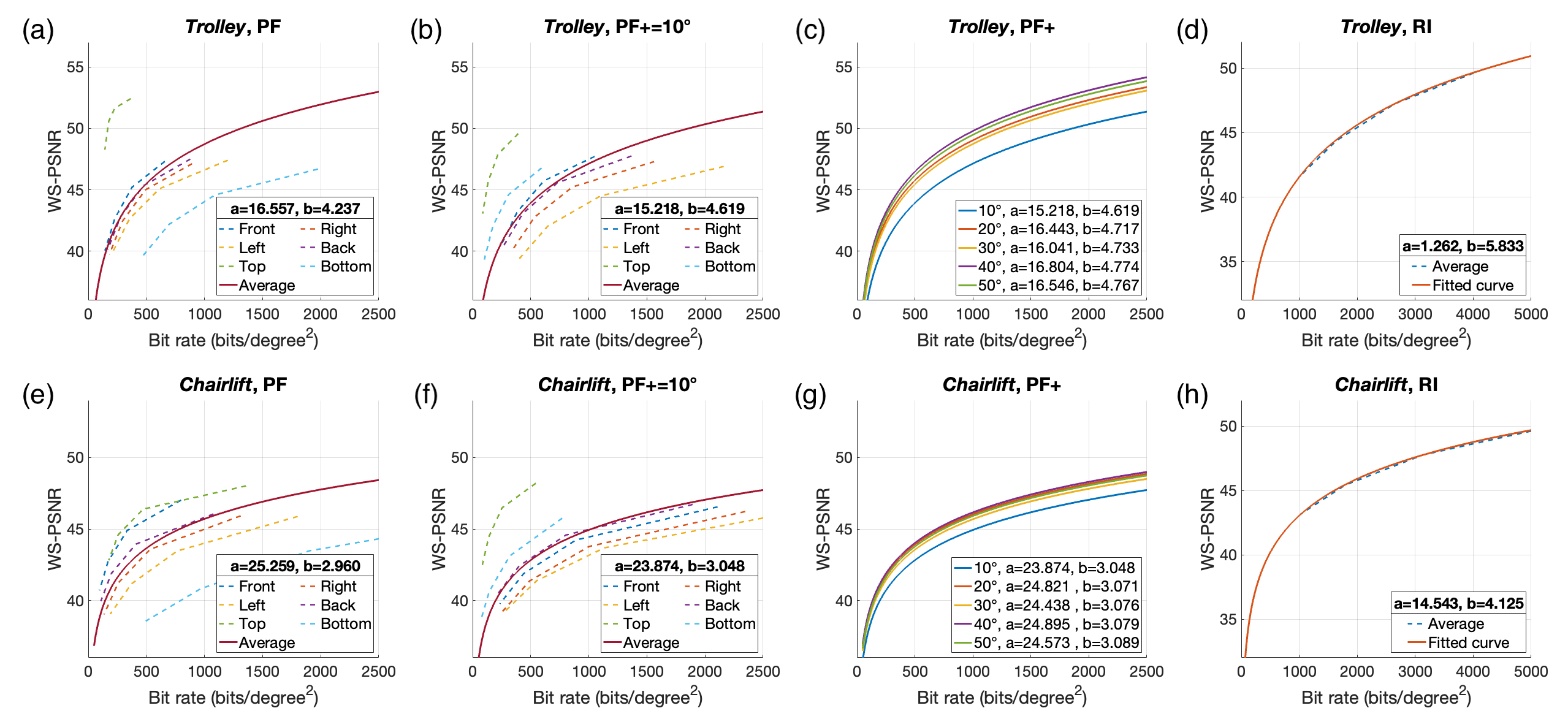}
    \vspace{-2mm}
    \caption{Q-R models for ``Trolley'' (a)-(d) and ``Chairlift'' (e)-(h). (a)(e): WS-PSNR vs. normalized rate for the PF regions of six viewing orientations, and the averaged WS-PSNR vs. normalized rate. (b)(f): WS-PSNR vs. normalized rate for the PF+ regions when its size is $10^{\circ}$. (c)(g): the averaged WS-PSNR vs. normalized rate for different PF+ region sizes.  (d)(h) WS-PSNR vs. normalized rate for the RI region.}
    % when the PF+ region sizes are $10^{\circ}$, $20^{\circ}$, $30^{\circ}$, $40^{\circ}$, $50^{\circ}$, respectively
    \label{fig:qr}
    \vspace{-5mm}
\end{figure*}

The RI region is coded using the intra-mode. The quality is the average WS-PSNR of all pixels in a RI, while the rate is the total bits of all pixels in the RI normalized to the average spherical area represented by RIs in different locations on the ERP. Figure \ref{fig:qr}(d) and \ref{fig:qr}(h) show the average Q-R functions. Note that the Q-R function is the same regardless the RI size, because all the tiles in a RI are considered equally useful (with a probability to be viewed characterized by the hit rate of $\alpha_{I}$). Again, this curve can be approximated well using a logarithmic function:
 \begin{equation}
    Q_{\rm RI}(R) = a_{\rm RI} + b_{\rm RI} \log R .
 \end{equation}

%\cite{supplementary}.

% \textbf{To circumvent the variability of the relation between the spherical area to be covered and the actual number of tiles, which depends on the actual location of the FoV, we define the bit rate in terms of the total bits needed to cover a unit area on the sphere.} Hence, the unit of normalized bit in the coding experiment and Figure \ref{fig:qr} is bits/degree$^2$.
Note again that all normalized bit rate in the coding experiment and Fig.~\ref{fig:qr} is defined in terms of bits/degree$^2$, i.e. the number of bits needed to cover a unit area on the sphere (the shape of FoV projection on ERP depends on the actual location of the FoV).

\subsection{Rate-Increase Factor}
\label{sec:rho} 
The ``ideal" Q-R models described in Sec.~\ref{sec:qr} are based on the assumption that PF and PF+ never change their location over time, which is not true in actual $360^{\circ}$ video viewing behavior. 
When the FoV changes between frames, a tile of the PF (or PF+) region in the current frame may not be coded in the previous frame, or even past several frames, as shown in Fig.~\ref{fig:codeing_dependency}.
{\it The coding time lapse} $\tau$ is defined as  the frame distance from the frame that the tile was last coded to the current frame. Generally in inter-coding, the accuracy of the temporal prediction reduces when $\tau$ increases. In other words, more bits are required to encode the frame at the same quality when $\tau$ is larger.
% To reach the same reconstruction quality, a higher rate is likely needed than when $\tau=1$. 
Therefore, we define {\it rate-increase factor} $\rho(\tau)$ to be the ratio of the rate needed for a given $\tau$ over the rate for $\tau=1$ to  achieve the same quality. Note that generally $\rho(\tau)$ depends on the QP and  the video content under the same $\tau$.
  
\begin{figure}[t]
\centerline{\includegraphics[scale = 0.15]{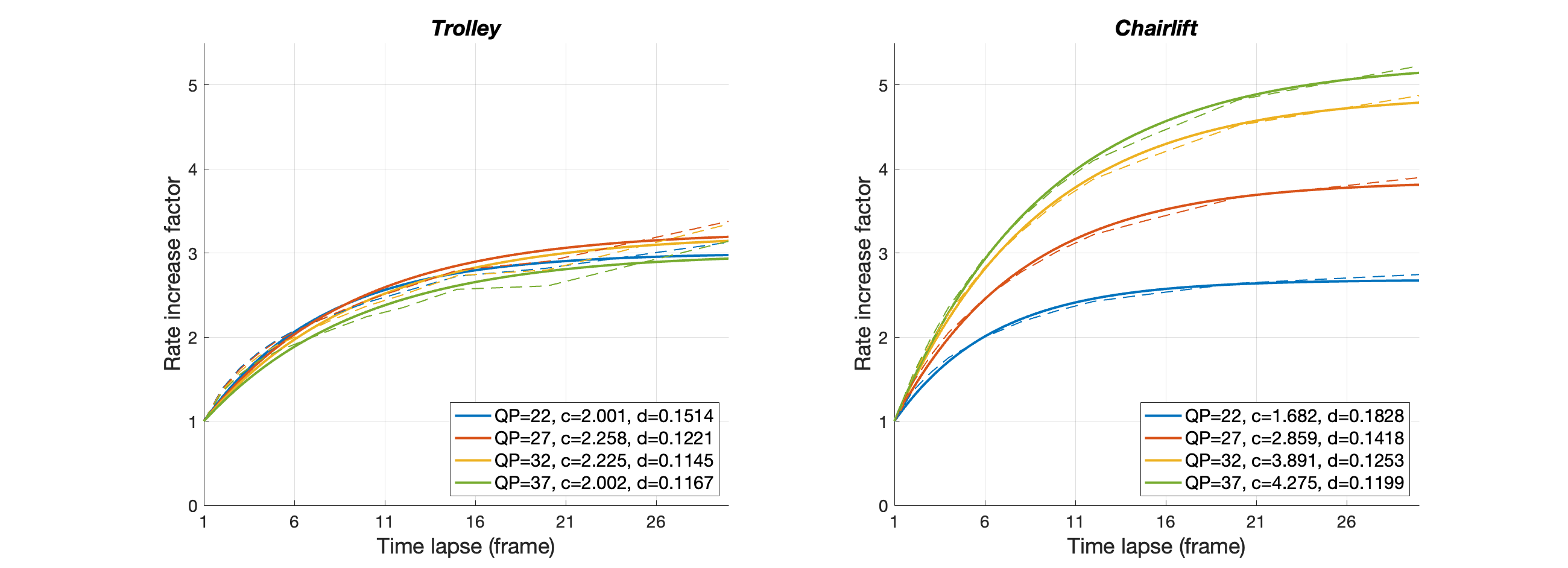}}
 \vspace{-2mm}
\caption{The rate-increase factor to maintain the same quality as a function of the time lapse between the inter-coded frame and the reference frame. Left: ``Trolley", fixed camera. Right: ``Chairlift", moving camera.}
\label{fig:rho}
\vspace{-4mm}
\end{figure}

In order to model $\rho(\tau)$, we code the testing videos using fixed QPs with different time lapses to measure the additional bits needed to achieve the same video quality. Fig.~\ref{fig:rho} shows the measured $\rho(\tau)$ for $QP = 22,27,32,37$ for two testing videos. As shown in the figure, the rate-increase factor can be well fitted by a reverse exponential decay function:
%\vspace{-3mm}
\begin{equation}
\label{eq:rate_increase}
    \rho(\tau) = 1+c \left( 1-e^{-d(\tau-1)} \right ).
%    \vspace{-1mm}
\end{equation}
The parameters $c$ and $d$ depend on the QP and the content. 

\begin{figure}
\centerline{\includegraphics[scale = 0.15]{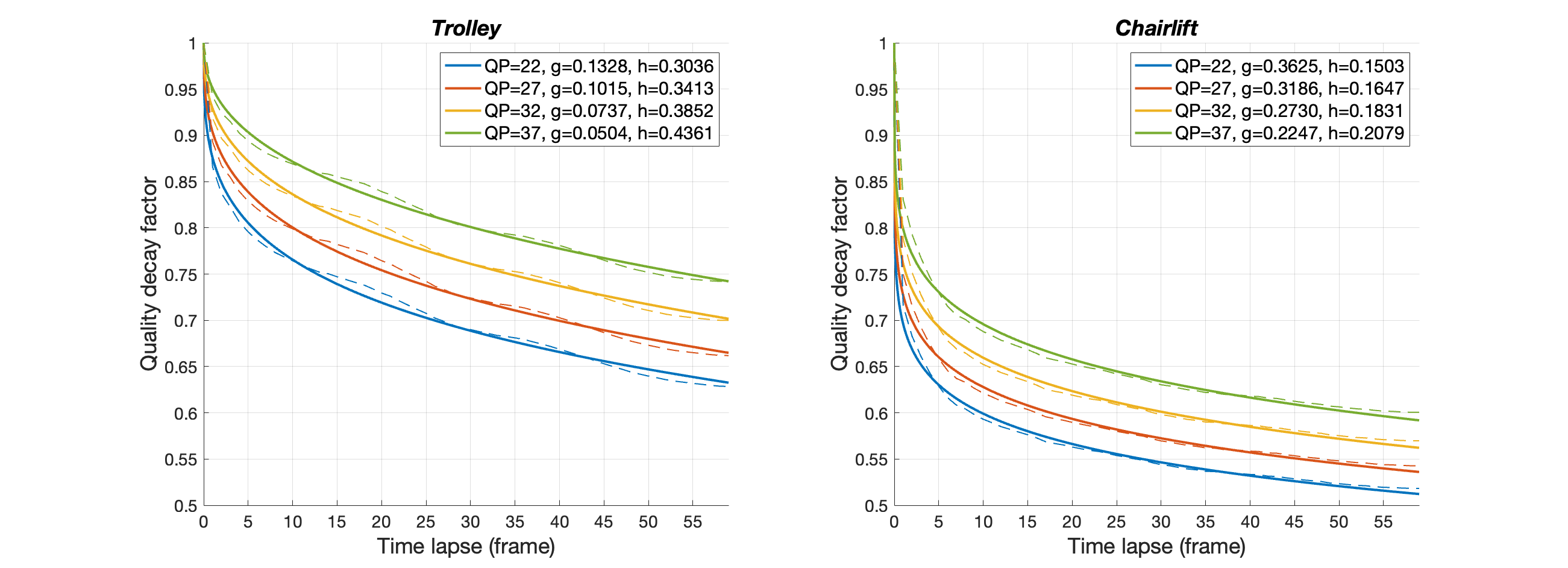}}
\caption{The quality-decay factor of pixels due to frame copy as a function of the time lapse between the last-coded frame and the current frame. Left: ``Trolley", fixed camera. Right: ``Chairlift", moving camera.
}
\label{fig:kappa}
\end{figure}

To achieve a given quality $Q$, the rate needed when $\tau=1$ is given by the ``ideal'' rate $R(Q)$ determined using the ``ideal'' Q-R models of the PF or PF+ region. The rate corresponding to other $\tau$ is given by 
%\vspace{-1mm}
\begin{equation}
    {\tilde R}(Q;\tau)= \rho(\tau) R(Q).
\end{equation}

\subsection{Adjust Quality-Rate Functions for PF and PF+ regions}

% In reality, the time lapse of tiles are temporally and spatially variant due to the FoV dynamics.
% To adapt the coding rates and the region sizes at the beginning of each video segment, we adjust the Q-R functions derived in Sec.~\ref{sec:qr} for the PF and PF+ regions based on the distribution of $\tau$ in the previous segment 

Given a QP, the actual rate required to code a tile inside the PF or PF+ region depends on the time lapse since this tile was last coded. The time lapse of each tile is spatially and temporal variant and depends on the FoV dynamics. To adapt the coding rates and the region sizes at the beginning of each segment, we adjust the Q-R functions derived in Section \ref{sec:qr_pf} and \ref{sec:qr_pf+} as follows:

{\parindent=0pt
\begin{enumerate}
\item Calculate the $\tau$ distribution of tiles in PF regions in the previous segment.
\item Locate the quality and rate values for each of the 4 QP values on the original average Q-R functions in Fig. \ref{fig:qr}.
\item For the rate in each sample point, calculate the rate-increase factor for each $\tau$ value using equation (\ref{eq:rate_increase}) and hence the adjusted rate. Then determine the average rate among all possible $\tau$'s based on the distribution of $\tau$. This will form a new Q-R point, where Q is the same as before, but R increased.
\item Use the new Q-R points to fit a new average Q-R function as follows:
%\vspace{-3mm}
\begin{equation}
    {\tilde R}(Q)=\left ( \sum_{\tau} p(\tau) \rho(\tau) \right ) R(Q),
    \label{eq:average-rate-increase}
\end{equation}
where $p(\tau)$ is the probability of $\tau$ measured among all the tiles in the PF region in the previous segment.
\end{enumerate}
}

The adjustment for the Q-R functions of PF+ region with variable region sizes can be done similarly. 
% Note that the distribution of $\tau$ depends on the FoV dynamics. When the FoV is fairly stationary, most coded tiles have $\tau$ close to one, and the adjusted Q-R functions are similar to the original ones. Because the FoV dynamics is time varying, we use the $\tau$ distribution determined from the predicted FoV trace in the last video segment to adjust the Q-R functions to be used for rate-allocation optimization for the new segment. 

\subsection{Quality-Decay Factor}
\label{sec:kappa}
As shown in Fig.~\ref{fig:codeing_dependency}, the PF, PF+, and RI regions may not cover the entire actual FoV. 
For a tile in the actual FoV that is not be coded and updated, it will remain the same as when this tile was last coded.  The rendered quality depends on how long ago (time lapse $\tau$) when it was last coded, and also the quality of the last coded tile.
We define {\it quality-decay factor} $\kappa(\tau)$ as the ratio of the quality of such a tile after time lapse $\tau$ over the quality of the last coded tile. 
%The longer the time lapse since this tile was last coded, the lower this ratio. 

In our experiments, we estimate the decay factor through simulation. For a given video sequence and a given QP, for each frame, we use the WS-PSNR derived from the ``ideal" experiment described in Sec.~\ref{sec:qr} as the WS-PSNR with $\tau=0$, denoted as $\mbox{WS-PSNR}(0)$. For the following $\tau$-th frame, we calculate the WS-PSNR between the coded first frame and the raw $\tau$-th frame, represented by $ \mbox{WS-PSNR}(\tau)$. The quality-decay factor is defined by $\kappa(\tau) = \mbox{WS-PSNR} (\tau) / \mbox{WS-PSNR} (0)$. We repeat this experiment using each of the first 200 frames in each of the two videos as the initial frame, and use the average decay factors for all the 200 samples as the decay factor for the given $\tau$. We repeat this process for $\tau$ between 1 and 100 to determine the $\kappa (\tau)$ function. As shown in Fig.~\ref{fig:kappa}, the quality-decay factor can be well fitted by a modified exponential decay model:
%\vspace{-2mm}
\begin{equation}
    \kappa(\tau) = e^{-g \tau^h}.
 %   \vspace{-2mm}
\end{equation}
Note that the values of $g$ and $h$ also depend on the video content and QP.
For a given rate, if the quality of the tile when it was last coded is $Q(R)$, the quality of the rendered tile that is $\tau$ frames away is determined by
\begin{equation}
{\tilde Q}(R; \tau)= \kappa(\tau) Q(R) .
\end{equation}

\subsection{Optimizing Rate Allocation and Region Sizes}
\label{sec:rate_size_allocation}
%\vspace{-1mm}
\subsubsection{Expected video quality}

% \textbf{With our coding scheme, the perceived quality of a rendered pixel depends on whether this pixel is coded in the PF, PF+, RI or is uncoded.}
The perceived quality of a rendered pixel depends on the coding region it falls in. Let
$\alpha_{\rm PF}$ denote the probability that a rendered pixel is in the PF region without overlapping with the RI region, to be called the {\it hit rate} of the PF region.
%\footnote{Because the RI is rolling, certain part of the PF may be covered by the RI in some frames.}  
Similarly, $\alpha_{\rm PF+}$ and $\alpha_{\rm RI}$ denote the probabilities that a rendered pixel is in the PF+ region (without overlapping with RI) and the RI region, respectively.  %Referring back to Figure \ref{fig:frame_structure}, the hit rate of each region is the percentage of pixels in the actual FoV that are in each region. 
Obviously $\alpha_{\rm PF}$, $\alpha_{\rm PF+}$, and $\alpha_{\rm RI}$ depend on the accuracy of FoV prediction. $\alpha_{\rm PF+}$ and $\alpha_{\rm RI}$ also depend on the sizes of the PF+ and the RI regions.

When a pixel in the actual FoV falls in the PF, PF+, or RI region, with the probability $\alpha_{\rm PF}$, $\alpha_{\rm PF+}$, and $\alpha_{\rm RI}$, respectively, it is decoded with quality $Q_{\rm PF}(R_e)$, $Q_{\rm PF+}(R_b)$ and $Q_{\rm RI}(R_b)$, correspondingly. 
For pixels not covered by these regions, they repeat the content last decoded. The quality for such a pixel is denoted as $\kappa(\tau) Q_{\rm last}$, where $\tau$ is the time lapse (frame distance) since it was last updated and $Q_{\rm last}$ is quality when last coded, as explained in Sec.~\ref{sec:kappa}. 
Generally, $\tau$ is varying in both space and time, and $Q_{\rm last}$ can be either $Q_{\rm PF}(R_e)$, $Q_{\rm PF+}(R_b)$ or $Q_{\rm RI}(R_b)$. 
In practice, since the rendered pixels have very small chance to fall in the un-coded region (lower than $1\%$  in our simulations), we can use the worst-case $\kappa_{\min} Q_{\rm RI}(R_b)$ to conservatively estimate its quality, where $\kappa_{\min} = \kappa(\tau_{\max})$,  with $\tau_{\max}$ being the full-ERP intra-refresh time (inversely proportional to the RI size).
Hence, the rendering quality of the actual FoV can be written as
%\vspace{-2mm}
\begin{equation}
\begin{split}
Q_1= &\alpha_{\rm PF} Q_{\rm PF}  +  \alpha_{\rm PF+} Q_{\rm PF+}  + \alpha_{\rm RI} Q_{\rm RI} \\
&+ (1-\alpha_{\rm PF} - \alpha_{\rm PF+} - \alpha_{\rm RI} ) \kappa_{\min} Q_{\rm RI}.
%\vspace{-2mm}
\end{split}
\end{equation}
% where $\lambda$ is fraction of pixels inside RI that are not covered by PF or PF+. 

$Q_1$ is the quality at the receiver when all the coded bits for this frame arrive  in time. When the bits for a frame arrive later than its display deadline,  the previously decoded frame is simply repeated and we can conservatively estimate the average quality as $Q_2 = \kappa_{\min} Q_{\rm RI}$. %Generally, the time lapse due to frame delivery failure is shorter than the average refresh period of the RI. But for simplicity, we will assume the decay factor is the same. 
Let $\gamma$ denote the {\it frame delivery rate}, which is the probability of in-time delivery.
The overall expected quality can be expressed as 
% \vspace{-2mm}
\begin{equation}
\label{eq:quality}
\begin{split}
&\bar{Q}(R_b, R_e,  A_{\rm PF+}, A_{\rm RI}) = \gamma Q_1 + (1-\gamma) Q_2 \\
&= \gamma (\alpha_{\rm PF} Q_{\rm PF}(R_e) + \alpha_{\rm PF+}Q_{\rm PF+}(R_b)+\alpha_{\rm RI} Q_{\rm RI}(R_b)) \\
&+ (1-\gamma(\alpha_{\rm PF}+\alpha_{\rm PF+}+\alpha_{\rm RI}))\kappa_{\min} Q_{\rm RI}(R_b),
%\vspace{-2mm}
\end{split}
\end{equation}
%Note that this quality definition does not consider the impact of quality variation within FoV in the same frame, nor the quality variation between adjacent frames. In our experimental results, we will report this  average quality, as well as temporal quality variation and spatial quality variation. 
%In Section \ref{sec:qr}, we discuss how to empirically determine the Q-R functions of different regions. 
where $A_{\rm PF}$, $A_{\rm PF+}$, and $A_{\rm RI}$ are the sizes of the PF, PF+, and RI region (in unit of the square degree), respectively. 
Note that $\alpha_{\rm PF+}$ and $Q_{\rm PF+}(R)$ depend on $A_{\rm PF+}$, and $\alpha_{\rm RI}$ is determined by $A_{\rm RI}$. Therefore, Eq.~(\ref{eq:quality}) is a function of $A_{\rm PF+}$, $A_{\rm RI}$, $R_e$, and $R_b$, for a given FoV prediction accuracy characterized by $\alpha_{\rm PF}$, and the frame delivery rate $\gamma$.

\subsubsection{Formulation and Solution of the Optimization Problem}
\label{sec:optimization}

Given the target bit budget $B_t$ of a frame, the region sizes $A_{\rm PF}$, $A_{\rm PF+}$, and $A_{\rm RI}$ and the corresponding normalized rates $R_b$ and $R_e$ need to satisfy:
 %\vspace{-1mm}
 \begin{equation}
 \label{eq:bwconstraint}
      \lambda_{\rm PF}A_{\rm PF}R_e + (\lambda_{\rm PF+} A_{\rm PF+}+ A_{\rm RI})R_b \leq B_t,
  %     \vspace{-1mm}
 \end{equation}
where $\lambda_{\rm PF}$ is the average ratio of tiles in the PF and not covered by the RI region, and $\lambda_{\rm PF+}$ is the same for the PF+. 
Both ratios are estimated by dividing the number of the RI tiles by the number of all tiles on the ERP frame.

%Note that in the actual coding process, we need to determine the total  number of tiles needed to cover each region based on the predicted FoV center, and consequently to derive the bits per tile or bits per pixel on the ERP for each region.

Since the PF size is fixed, the goal is to find the optimal region size combination ($A_{\rm PF+}$, $A_{\rm RI}$) and corresponding rates ($R_b$ and $R_e$) to maximize the quality shown in Eq.~(\ref{eq:quality}) subject to the target bit budget constraint in Eq.~(\ref{eq:bwconstraint}). 
Generally, $\alpha_{\rm PF+}$ and $\alpha_{\rm RI}$ increase with larger $A_{\rm PF+}$ and $A_{\rm RI}$, and $\kappa_{\min}$ also increases with larger $A_{\rm RI}$. However, the rates $R_e$ and $R_b$ decrease with larger $A_{\rm PF+}$ and $A_{\rm RI}$ due to the target bit budget constraint. 

To simplify the practical system setting, we limit the possible sizes of the PF+ and the RI within a finite candidate set.
% This system was in my cart 2 days at $1100 with all coupon applied. 
For each possible combination of the PF+ and RI size, only $R_e$ and $R_b$ in Eq.~(\ref{eq:quality}) are the free variables. Given that the optimal solution lies when the bit budget is met exactly in Eq.~(\ref{eq:bwconstraint}), we have 
$R_b= (B_t-\lambda_{\rm PF}A_{\rm PF}R_e)/(\lambda_{\rm PF+}A_{\rm PF+}+ A_{\rm RI})$. 
Then, the optimal $R_e$ can be derived by setting $\frac{\partial \bar{Q}}{\partial R_e}= 0$. We apply the log Q-R model introduced in Sec.~\ref{sec:qr} and get the analytical solution as:
%\vspace{-1mm}
\begin{equation}
R_e = \frac{X}{X+Y}\frac{B_t}{\lambda_{\rm PF}A_{\rm PF}},  
\; 
R_b = \frac{Y}{X+Y}\frac{B_t}{\lambda_{\rm PF+}A_{\rm PF+} + A_{\rm RI}},
\end{equation}
where
%\vspace{-4mm}
\begin{eqnarray*}
X&=&\gamma\alpha_{\rm PF}b_{\rm PF} ,\\
Y&=&\gamma\alpha_{\rm PF+}b_{\rm PF+}+\gamma\alpha_{\rm RI}b_{\rm RI}+\kappa_{\min} b_{\rm RI} \\
   & & -\gamma\kappa_{\min} b_{\rm RI}(\alpha_{\rm PF}+\alpha_{\rm PF+}+\alpha_{\rm RI}) .
\end{eqnarray*}
We enumerate all possible region sizes and the corresponding optimal rate combinations to find the optimal combination maximizing $\bar{Q}$.

%------------------------------------------------------------------------- 
\section{FoV and Bandwith Prediction}
\label{sec:fov_bd_section}
In this section, we describe the methods used for frame-level FoV prediction and segment-level bandwidth prediction.

\begin{figure}
\centerline{\includegraphics[scale=0.155]{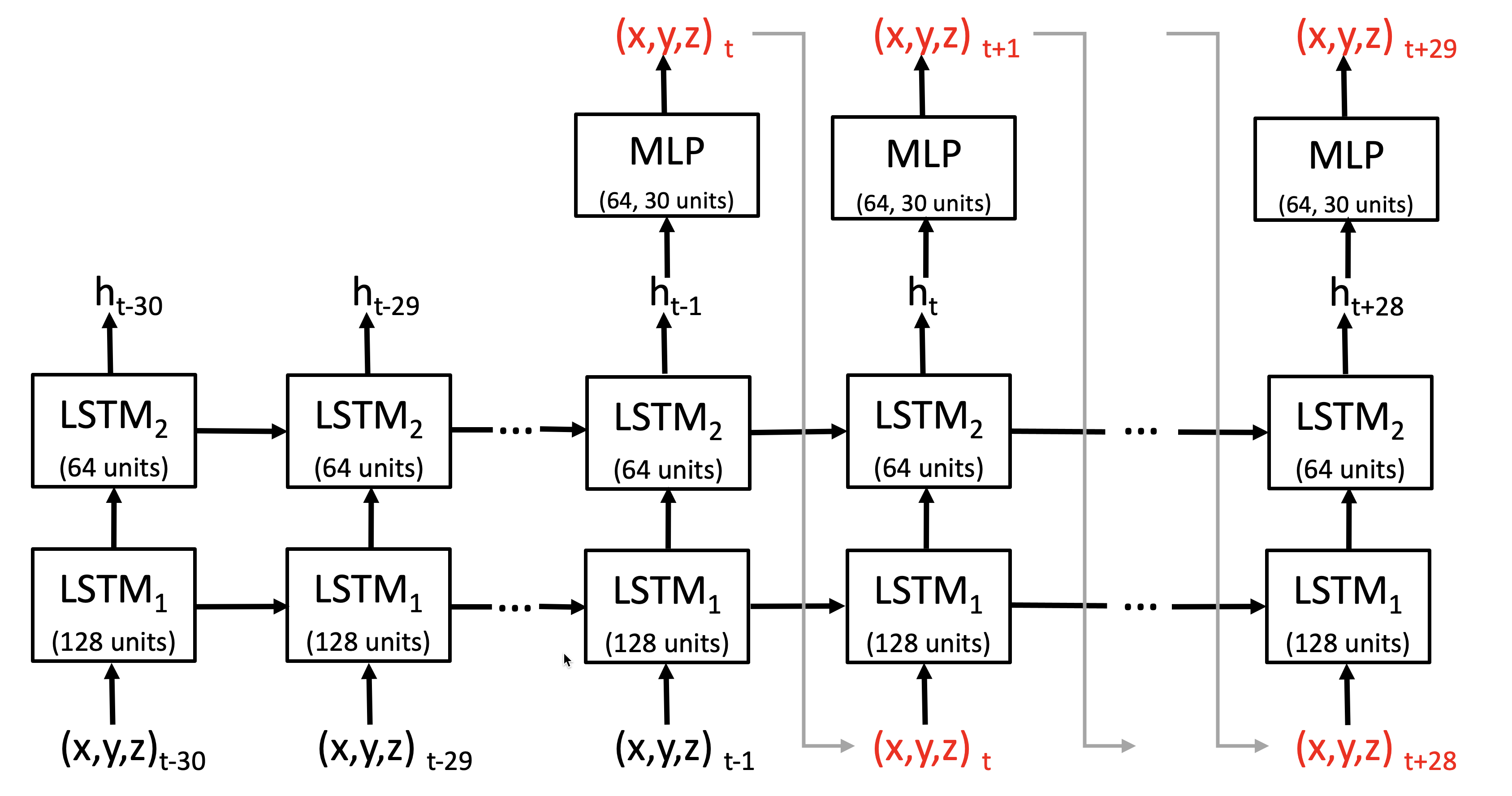}}
% \vspace{-2mm}
\caption{The LSTM model for FoV prediction, numbers of hidden units are indicated in layer blocks.}
\label{fig:fov_lstm_model}
% \vspace{-6mm}
\end{figure}

\begin{figure}
\centerline{\includegraphics[scale=0.15]{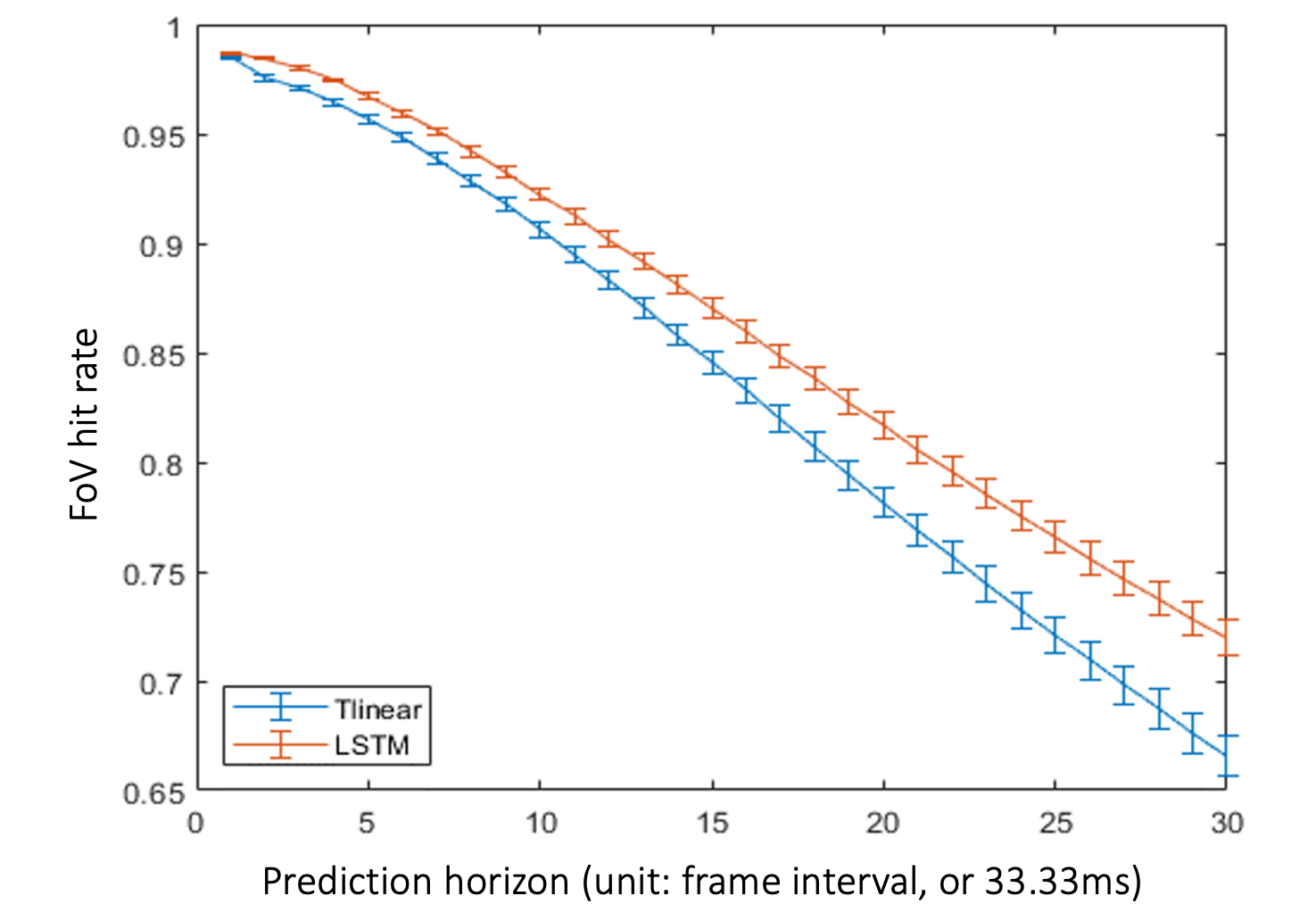}}
% \vspace{-2mm}
\caption{Hit rate of predicted FoVs, using the LSTM model and the Truncated Linear predictor, on the testing data. Error bar: 3$\times$ standard error of the mean (SEM).
}
\label{fig:fov_lstm}
\vspace{-4mm}
\end{figure}

\subsection{Frame-Level FoV Prediction}
\label{sec:FoV-prediction}
%\textbf{ Newly add: }FoV prediction is critical to the performance of FoV-adaptive streaming. Most of solutions  predict the future FoV center from the FoV centers in the past.  For example,  linear regression, weighted linear regression, and truncated linear prediction are proposed in \cite{fov_adapt_2,qian2016optimizing,1}. More sophisticated network-based methods are proposed in \cite{fov_pred,pred1,pred2}. Most of these methods  can predict the short-term FoV (within the future 1 second) well with an accuracy of more than 90\% \cite{qian2016optimizing}\cite{fov_pred}\cite{pred2}. %Yet, the FoV prediction accuracy drops significantly for long term even for the state-to-art deep learning predictor. 
The performance of FoV-adaptive $360^{\circ}$ video streaming highly depends on the FoV prediction accuracy. Multiple time series prediction methods have been applied on this topic in previous works, e.g. linear regression, weighted linear regression, truncated linear prediction \cite{fov_adapt_2,qian2016optimizing,1}, and deep-learning (DL) methods \cite{fov_pred,pred1,pred2,8779683}. 
Although most methods predict the short-term FoV (within the future 1 second) well with the accuracy of more than 90\% \cite{qian2016optimizing}\cite{pred2}, DL-based methods still outperform conventional methods especially when the prediction horizon is long \cite{fov_pred}.

We use the popular long short-term memory (LSTM) architecture, which is one of the most suitable neural networks to make predictions based on time series data. The input to the LSTM model consists of the FoV center locations described by Cartesian coordinate $(x,y,z)$ over the past 30 frames. Note that we choose not to use the $(yaw, pitch)$ angles to avoid the issue of $2\pi$ periodicity of $yaw$. The hidden states corresponding to each future frame are mapped to the predicted FoV center locations through two fully connected layers. The predicted location for each new frame is recursively fed to the input for the next frame time, until the desired prediction horizon is reached. We experimented with LSTM models with single, two, and three fully connected LSTM layers. We find the networks with two or three LSTM layers achieve similar prediction accuracy in terms of the FoV hit rate, while they both outperform the single layer model. Therefore, we adopt a model with two LSTM layers and the model structure is shown in Fig.~\ref{fig:fov_lstm_model}. The two LSTM layers have 128 and 64 hidden units, and the two fully connected (FC) layers contain 64 and 30 hidden units, respectively.
% (Yixiang: add this figure, indicate the hidden unit numbers in the figure.)
This simple structure provides sufficiently accurate results for the short prediction horizon of interests (typically under 300 ms or 10 frames), while enjoying relatively low computational complexity.

We use the FoV hit rate to evaluate the prediction accuracy, which is defined as the overlapping ratio of the predicted FoV and the ground truth FoV on the unit sphere.
\textcolor{black}{We train our model using the FoV trace data from \cite{shanghai}. We choose 20\% of the traces as the testing set (including the traces used in the system simulation). Then we split the remaining traces  into a training set (80\%) and a validation set (20\%). We choose the model's hyper-parameters to maximize the FoV hit rate on the validation set.}
Figure~\ref{fig:fov_lstm} shows the FoV hit rate on the test set. Compared to the truncated linear prediction method used in our preliminary study \cite{1}, which uses the last few past samples among a maximum number of past samples that can be approximated well by a linear function to predict a future sample, the LSTM model is significantly more accurate.  
% As shown in Figure ~\ref{fig:trace}, this method can generate very accurate prediction except when the FoV suddenly changes. Furthermore, the predicted trace has a very short time lag from the actual trace after sudden changes.

% \begin{figure}
% \centerline{\includegraphics[scale=0.22]{TMM/LSTM_overall.png}}
% % \vspace{-2mm}
% \caption{Illustration of the overall bandwidth prediction model.}
% \label{fig:lstm}
% % \vspace{-6mm}
% \end{figure}

% \begin{table}
% \centering
% \setlength\tabcolsep{4.5pt}
% \begin{tabular}{|l|c|c|c|c|c|c|} 
% \hline
%     & RLS & Model 1 & Model 2 & Model 3 & Combined & +history  \\ 
% \hline
% MRAE & 24.8\% & 24.2\% & 24.4\% & 25.4\% & 21.2\% & 20.6\% \\
% \hline
% \end{tabular}

% \caption{Mean relative absolute error (MRAE) of different bandwidth prediction algorithms.}
% \label{tab:bw_pred}
% \end{table}

\subsection{Segment-Level Bandwidth Prediction}
\label{sec:bandwidth-prediction}
Bandwidth prediction is critical to the performance of rate-adaptive streaming systems. Many  methods have been proposed to predict the network bandwidth in prior works, including Harmonic Mean\cite{jiang2012improving}, Recursive least square (RLS)\cite{eymen}, Random Forest\cite{yue2017linkforecast}, and Hidden Markov Model\cite{sun2016cs2p}. More recently, deep learning models (including LSTM-based) have shown advantages over prior methods \cite{mei2020realtime,band_pred,9154522}.

In our streaming system, we predict the average sustainable throughput from the sender to the receiver over the next segment time (1 sec.) at the beginning of the new segment, based on the measured throughput at the intervals of 200ms in the past three segments (3 sec.) returned by the receiver. 
% Inspired by the model selection method from a prior work \cite{band_pred}, we propose a fusion model that can effectively adapt to the change in the network dynamics. This is achieved by adaptively combining the predicted bandwidth by LSTM models driven by bandwidth histories of different lengths, shown in Fig.~\ref{fig:lstm}. The fusion model includes two parts. First, We train three LSTM models to predict the bandwidth of the future segment, using measured bandwidth with history windows of 1s, 3s, or 10s, respectively. 
We use a LSTM sequence-sequence model, with a structure very similar to that for FoV prediction shown in Fig.~\ref{fig:fov_lstm_model}, but with different numbers of hidden units in each layer. The bandwidth prediction model has two LSTM layers with 96 and 64 hidden units, followed by two FC layers with 64 and 5 hidden units, respectively.
% The structure of each model is shown in Figure \ref{fig:lstm_unfolded}, with different history window size $\tau$. 
Note that we use the average throughput over a 200ms window as the input feature at each time step, hence, the model has 15 input samples. The model recursively predicts the throughputs for the five consecutive 200ms windows in the next second. The sum of these five predicted throughputs is the predicted total throughput for the next second, $\tilde{B_t}$.

\textcolor{black}{We train our model using the LTE packet traces collected in \cite{eymen}. We choose a trace named ``att-downlink" as the testing trace and it is used in the following simulation experiment. The remaining traces are divided into overlapping 4 sec. long short sequences, and 80\% of the short sequences from each trace are used to form the training set, and the remaining 20\% are used for validation.  
We choose the model's hyper-parameters and input window length  (among 1 sec., 3 sec. and 5 sec.) based on the prediction errors on the validation set. The window length of 3 sec. was found to perform slightly better than the other choices.} 

% The second part is a multilayer perception (MLP) that is used to calculate the weights for combining the outputs from the three LSTM models. The input to the MLP model is the bandwidth history of the past 10 seconds. Intuitively, the MLP model should assign a larger weight to Model 1 when the bandwidth is  very dynamic, and assign a larger weight to Model 3 when the network is stable. Fig.~\ref{fig:lstm} shows the overall fusion model structure. 
%We have $B_T=\sum_{i=5T}^{5T+4}b_i$ since each segment contains five 200ms windows. 

% We also monitor the bandwidth prediction accuracy in the past segments. If the prediction error in the past segment is more than twice of the average error in the past, it means the recent network throughput is very dynamic and hard to predict. In this case, we force the system to use the bandwidth predicted by Model 1 which uses the shortest bandwidth history.
We compare the performance of our model with RLS \cite{eymen} using Mean Absolute Percentage Error (MAPE) and normalized Mean Absolute Error (nMAE), defined as 
\begin{equation}
    \mathrm{\mbox{MAPE}}= \frac{1}{T}\sum_t min \left (\frac{\left | \tilde{B_t}-B_t \right |}{B_t}, 1.0 \right),
\end{equation}
\vspace{-2mm}
\begin{equation}
    \mathrm{\mbox{nMAE}}=\frac{\sum_t \left | \tilde{B_t}-B_t \right |}{\sum_t B_t},
\end{equation}
where $B_t$ and $\tilde{B_t}$ are the actual and the predicted bandwidth at segment $t$, respectively. MAPE calculates the relative error at each segment and it is more meaningful for our segmentation-level rate-adaptive streaming. We cap the relative error to $1.0$ to prevent the large error resulting from when the actual bandwidth is very small to dominate the reported performance.  

Compared to RLS, the MAPE of the proposed model is reduced from $21.1\%$ to $18.9\%$ on our testing trace, while the nMAE of the proposed model is also dropped from $14.1\%$ to $13.7\%$. 
%Figure~\ref{fig:bd_cdf} shows the cumulative distribution functions (CDFs) of the MAPE for the RLS and the proposed model.

%\begin{figure}
%\centerline{\includegraphics[scale=0.4]{TMM/bd_pred_cdf.png}}
%% \vspace{-2mm}
%\caption{CDFs of Mean Absolute Percentage Error for different methods.}
%\label{fig:bd_cdf}
%% \vspace{-6mm}
%\end{figure}

% In a prior work \cite{eymen} on rate adaptation for video calls, an online linear adaptive filter called recursive least square (RLS) was found to achieve higher prediction accuracy than exponentially weighted moving average. Therefore, we adopt the RLS method in this work and use the measured bandwidth in previous two segments for the prediction. As shown in Figure ~\ref{fig:trace}, the prediction accuracy is very good with a short time lag after a sudden change.

%------------------------------------------------------------------------- 
\section{Rate and Region Size Adaptation for Interactive Streaming}
\label{sec:streaming_section}

\subsection{Proposed streaming system overview}

\begin{figure}
\centerline{\includegraphics[scale=0.135]{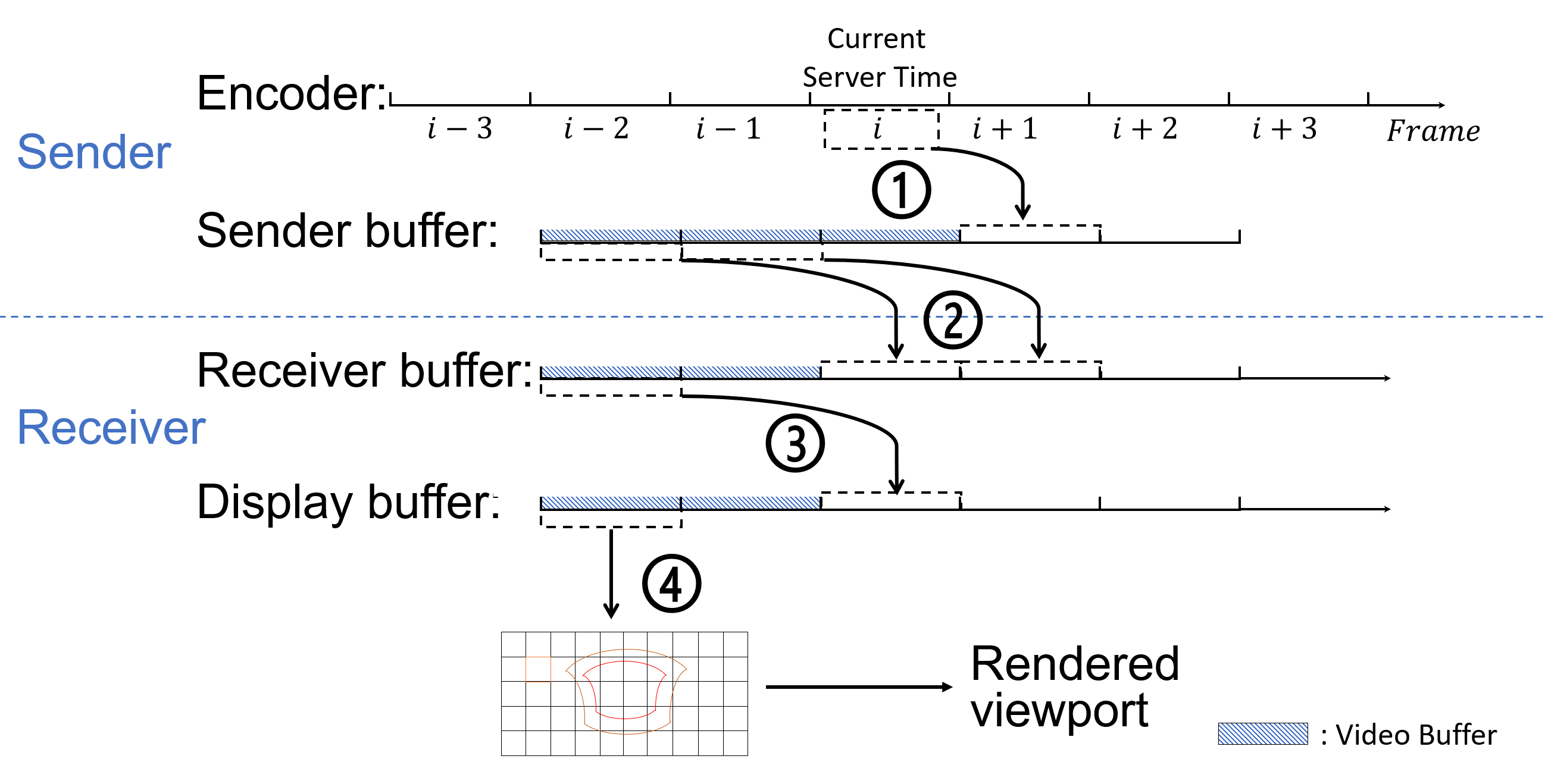}}
\caption{The proposed streaming system.}
\label{fig:streaming_system}
\vspace{-4mm}
\end{figure}

The proposed  $360^{\circ}$ video interactive streaming system uses the ``server push"  solution, where the server (or Sender as in Fig.~\ref{fig:streaming_system}) controls the schedule of video coding and packet sending.
The system predicts the network bandwidth ($\tilde B_t$) and region hit rates ($\alpha_{\rm PF}, \alpha_{\rm PF+}, \alpha_{\rm RI}$) for each segment at the beginning of encoding the segment (each segment is 1 second long including 30 frames in our experiments), based on the network throughput and the FoV history continuously fed-back by the receiver. Using the estimated bandwidth and region hit rates, the system performs the optimization described in Sec.~\ref{sec:rate_size_allocation} to calculate the sizes and average rates of the RI and PF+ regions for this segment.
The video frames in the segment are sequentially coded. The bit stream for each encoded frame is appended to the end of the sender buffer if the buffer is not full, as indicated by Process 1 in Fig.~\ref{fig:streaming_system}. If the sender buffer reaches its maximum capacity $B_{\max}$, this newly encoded frame will be dropped to reduce the accumulated delay. We set $B_{\max}=10$ frames in simulations. 
The server keeps pushing out as many frames as possible in the sender buffer to fully utilize  the available bandwidth, as shown in Process 2 in Fig.~\ref{fig:streaming_system}. 

Each newly received frame is decoded using the current reference frame in the receiver and appended to the end of the display buffer, as indicated by Process 3 in Fig.~\ref{fig:streaming_system}. 
The reference frame on the receiver will be updated to this newly decoded frame. Even if a frame arrives later than its display deadline, the receiver still decodes it to update the reference frame to avoid any possible mismatch with the encoder. 

The display checks the front of the display buffer every 1/3 frame interval. If the next decoded frame exceeds the maximum display delay (20 frames in our experiments), it will be dropped and the display checks the next frame in the display buffer until a frame meets the display deadline. The viewport will be rendered and displayed for each timely, shown as Process 4 in Fig.~\ref{fig:streaming_system}. If there is no frame in the display buffer or all frames in the buffer are too late to display, the last displayed frame will be repeated, leading to video freeze. 
%If several delayed frames arrive together, $3 \times$ playback speed is utilized to catch up, by checking the display buffer at a high frequency (one frame every 1/3 frame interval) (YM: This is the same as the 1/3 frame interval you mentioned before -- Yixiang: Then we only keep "If several ... to catch up"? or delete them?)
% Note that the frame freeze can be caused by either the skipped frame at the sender or the late frames at the receiver. 
Note that in our trace-driven simulations, we assume encoding each frame takes a constant 33.3ms and decoding a frame takes 11.1ms (33.3ms = 1 frame interval for the 30fps test videos). The reported frame delays in Table~\ref{tab:big_table} are already very low (average $<100$ms), and can be further shortened when a faster encoder or decoder is available.

\subsection{Adaptation of coding rates and region sizes}

We first determine the target bandwidth budget for the next segment based on the predicted available bandwidth for the segment and the current sending buffer occupancy. 
Then, we calculate the target sizes and bit rates of different regions (PF, PF+, and RI) for all frames in the next segment by maximizing the expected video quality of the segment formulated in Eq.~(\ref{eq:quality}), using the method described in Sec.~\ref{sec:optimization}. 
The system measures the average FoV hit rates of different regions and the frame delivery rate of the current segment and assume the FoV hit rate and frame delivery rate remain unchanged when solving  the optimization problem for the next segment. 
The system also calculates the time lapse distribution $p(\tau)$ in the current segment to determine the average rate increase factor using Eq.~(\ref{eq:average-rate-increase}) and correspondingly adjust the Q-R functions  derived in Section \ref{sec:qr-adjusted} for the next segment.

\subsubsection{Assigning the Total Bit Budget for a Segment Considering Sending Buffer Status} 
The packet can be occasionally backlogged in the sender buffer over time because of the error of the segment-level bandwidth prediction and the fluctuation of the actual network bandwidth within a segment. 
To avoid those packets from accumulating in the sending buffer, once we have the predicted bandwidth for next segment $s$, obtained using the bandwidth prediction method described in Sec.~\ref{sec:bandwidth-prediction},  we calculate the target bit budget by subtracting the bits $q_s$ currently left in the sender buffer from the predicted bit budget of the segment $\hat{b}_s$. 
Moreover, the bandwidth utilization ratio $\eta$ is applied to further lower the probability of exceeding the actual network capacity. 
Experimental work on real LTE traces shows that this probability can be kept lower than $5\%$ by setting $\eta<=66\%$\cite{eymen}. The target bit budget to encode segment $s$ is 
\begin{equation}
    b_s = \eta (\hat{b}_s - q_s).
\end{equation}

\subsubsection{Frame-level Bit Budget Update} 
\label{sec:frame_rate}
The bandwidth can be unstable within a segment, especially over the LTE or 5G wireless connection. The  bandwidth prediction model in Sec.~\ref{sec:bandwidth-prediction} only predicts the total bit budget inside a segment, so a more detailed adjustment at the frame level is necessary. The system adapts this frame-level bit budget by checking the space left in the buffer and the remaining bit budget of the segment. Each segment contains $N$ frames ($N=30$ in our simulation), on average $\frac{n}{N}b_s$ bits should have been used at the time of coding the $n$-th frame in the segment ($n=0$ for the first frame). However, the actual bits already used $S(n)$ in this segment when coding the $n$-th frame could differ from this average. The streaming strategy is designed to be conservative to reduce the risk of freeze and high delay, by setting the  remaining bit budget in the segment when coding the $n$-th frame as
\begin{equation}
\label{eq:regment_budget_adjust}
    b_s(n) = b_s - max\left (S(n), \frac{n}{N}b_s \right).
\end{equation}
The system further adjusts the rate of the $n$-th frame based on the sender buffer occupation $B(n)$. If the sender buffer is full ($B(n)=B_{\max}$), this frame would not be coded or transmitted. If the buffer is nearly full,  the target rate should be reduced from the average rate $\frac{b_s(n)}{N-n} $. The  target rate to code the $n$-th frame when the buffer is not full is determined by
\begin{equation}
    B_t(n) = \frac{b_s(n)}{N-n} a \exp(-b B(n)/B_{\max}),
\end{equation}
where $a$ and $b$ are parameters that can be adjusted empirically. We choose $a=1.20$, $b=1.00$, and $B_{\max} = 10$ frames in our simulations.

%------------------------------------------------------------------------- 
\section{Simulation Results}
\label{sec:simulation_exp}

\subsection{Test sequences, bandwidth and FoV traces}
\begin{figure*}[t]
\vspace{-4mm}
 \centerline{\includegraphics[scale = 0.3]{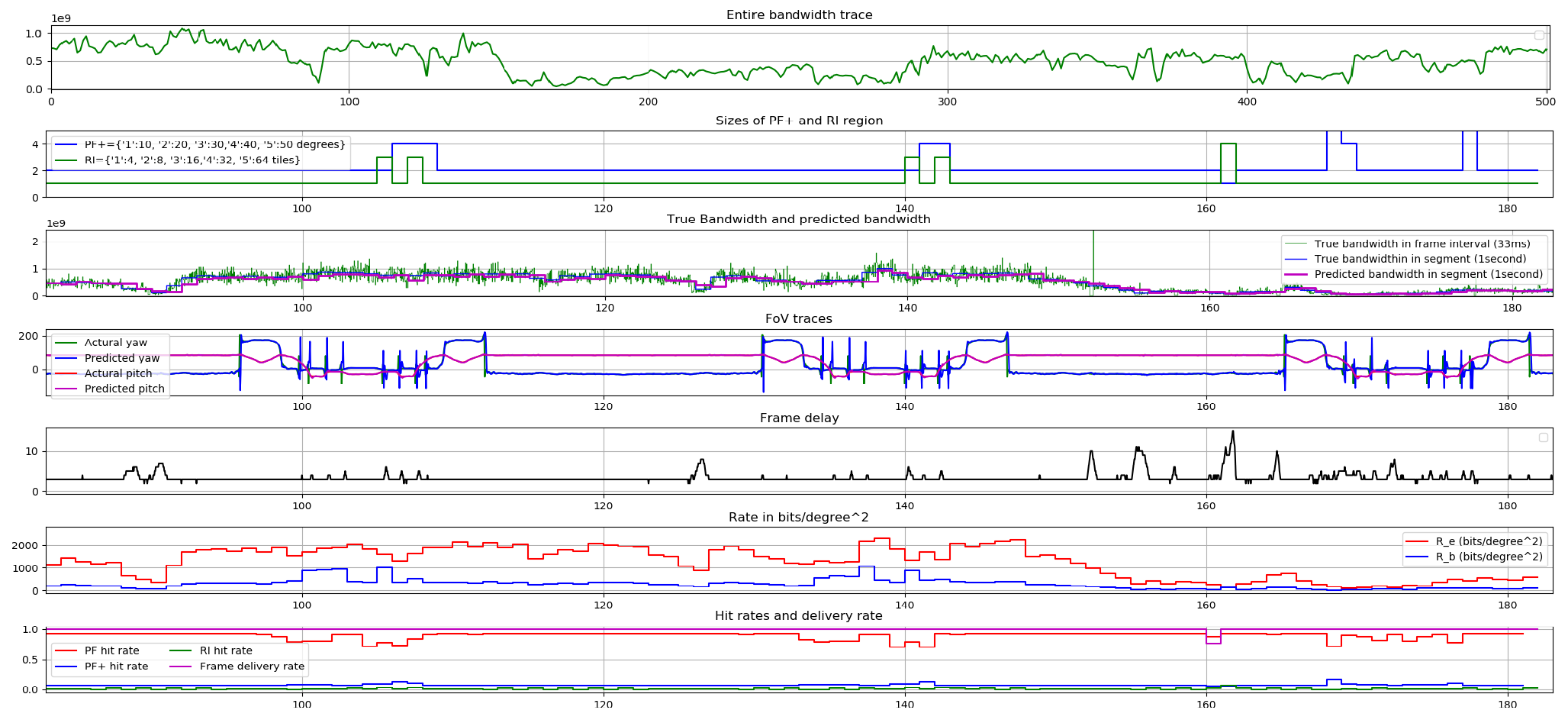}}
 \vspace{-3mm}
 \caption{The top plot: the LTE bandwidth trace collected in real world (500 seconds long); Other plots: various performance indices within a short time duration (90 second - 190 second). Horizontal axis is in unit of second.}
\label{fig:trace}
\vspace{-3mm}
\end{figure*}

We performed trace-driven simulations to evaluate the proposed coding and streaming system using real viewers' FoV traces and LTE network bandwidth traces. The LTE bandwidth traces are derived from the packet arrival time sequences collected in the real world as described in  \cite{eymen}.
To challenge our system, we run the simulation on a dynamic network trace (500 sec. long) with bandwidth variance over mean ratio ${\rm std/mean} = 0.673$, as shown in Fig.~\ref{fig:trace}. This trace includes periods where the bandwidth is high, low, and has sudden drops. To match the rate range for the 8K testing video, we scale up the range of our LTE bandwidth traces to have an upper-bound of 200 Mbps, which is realistic under future 5G networks. We set the one-way propagation time to 15ms in our simulations, which is typical for the network delay within the US \cite{networkdelay}.

We use two JVET $360^\circ$ test sequences in 8K ERP format,  ``Trolley'' and ``Chairlift'' \cite{wspsnr}, to evaluate the performance of our proposed and other benchmark streaming systems. ``Trolley'' contains a stable scene where the background is stationary, while ``Chairlift'' shows a more dynamic scene with a dynamic background. Each sequence has 300 frames in YUV 4:2:0 format with the resolution of 8192$\times$4096 at 30 frames per second. The bit-depths of ``Trolley'' and ``Chairlift'' are 8 and 10 bits, respectively. 

% (``ideal'' Q-R, the rate-increase, and quality-decay models)
We assume the category of the video content (e.g., fast-changing scene shot by a moving camera, or stable scene captured by a fixed camera) can be determined before or at the beginning of a video streaming session, and the parameters for the Q-R models for different categories can be pre-determined. Using these predetermined parameters, the system can perform  rate and region size adaptation as introduced in Sec.~\ref{sec:qr-adjusted} and \ref{sec:rate_size_allocation}. To handle the situation where the video scene category changes dynamically within a streaming session, some automatic ways to periodically updating the scene category need to be developed. Furthermore, we use the Q-R models introduced in Sec.~\ref{sec:qr-adjusted} and \ref{sec:rate_size_allocation} to determine the  quality of each tile given the rate allocation, instead of doing the actual video coding and decoding.
For each frame, the simulation system updates a table recording the time lapse and the quality of each tile when the tile was last coded.
As introduced in Sec.~\ref{sec:rho}, the actual bit rate to inter-code a tile is increased from the target rate by the rate increase factor $\rho(\tau)$ based on the time lapse $\tau$. 
To calculate the WS-PSNR of each tile in the displayed FoV of each frame, we use the recorded time lapse and the quality when it is last coded to determine its current quality by using the quality decay factor $\kappa(\tau)$ introduced in Sec.~\ref{sec:kappa}. 

Since each JVET test sequence only has a duration of 10 seconds, it is not reasonable to collect viewers' FoV trace only for such a short time period (it will be highly affected by the default initial FoV).
Therefore, we choose two groups of representative traces from open-source FoV trace datasets \cite{shanghai,tsinghua} for $360^{\circ}$ videos. 
For the stable-scene video ``Trolley'', we choose the traces collected by \cite{shanghai} where participants watched a video shot by a fixed camera named ``Weekly Idol-Dancing''. 
For the dynamic-scene video ``Chairlift'',  we choose the traces collected by\cite{tsinghua} where participants watched a video captured by a moving camera named ``GoPro VR-Tahiti Surf''. 
To remove the random jitters in the raw collected traces,  we apply Kalman filtering to the raw traces  and use the smoothed traces in our simulations.
Since the bandwidth trace is longer than the FoV trace, we extend each  FoV trace by appending the temporally flipped FoV trace to itself repeatedly to match the length of the bandwidth trace. The reported results in Table~\ref{tab:big_table} are the average results from simulations using 48 users' FoV traces, each of which is repeated to a duration of 500 seconds.

% \textcolor{red}{(Yixiang: unless you explained the coding set up for the proposed system, including the tile size earlier, it needs to be added here. I think you need to move some of the supplementary material here as a subsection)}

\subsection{Streaming System Benchmarks}
We also simulated three tile-based state-of-art streaming systems as the benchmarks for comparison, where BM1 and BM2 use intra-coding for all frames, and BM3 uses inter-coding. 
BM1 follows the  coding strategy in \cite{interactive_gamming}, which intra-codes and sends non-overlapped vertical slices centered at the predicted FoV center in each frame. The vertical slices cover a $140^{\circ}\times 180^{\circ}$ region on the ERP map while the actual FoV is $90^{\circ}\times90^{\circ}$. Note that such vertical slices cannot fully cover the FoV when the FoV was facing the poles (up and down). 
BM2 uses the same tile size as our proposed system, but it codes all tiles in both PF and PF+ regions only using intra-coding. For BM2, the size of PF+ is fixed to cover a $50^{\circ}$ border around the PF. 
Finally, BM3 applies inter-coding with periodic intra-frames. Rather than using rotating intra regions in the proposed system, BM3 codes the entire ERP of the first frame in each segment as the I frame. The remaining frames in the segment are inter-coded in both PF and PF+ regions with the same rate, and the size of the PF+ region is also fixed to cover a $50^{\circ}$ border. 
We use the Q-R models derived for the RI region (use intra-coding) to determine the WS-PSNR of coded tiles for BM1, BM2, and the I frames in BM3 for a given rate. We apply the Q-R models derived for the PF region (use inter-coding) for the inter-coded regions in BM3.
For BM3, the ratio of  the I-frame bit rate and P-frame rate is assumed to be equal to the average ratio measured from the actual coding experiment over a range of  QP. The normalized rate for the P-frame is set so that the total bits for coding the I-frame and all P-frames in each segment  is equal or below the target rate budget. 
For a fair comparison, all benchmark systems share the same elements with our proposed system, including the same bandwidth and FoV prediction algorithms, the same segment- and frame-level bit rate adjustments, and the same quality-decay model to calculate the quality of the un-coded region in the displayed FoV.

\subsection{Evaluation metrics}
\label{sec:eval_matrics}
We evaluate each streaming system on each test video and report the average values of various metrics resulting from using the 48 extended FoV traces.
Those metrics include the average frame delay and delay standard deviation (STD), the freeze frequency and duration, and the average rendering quality (average WS-PSNR of all pixels in the actual viewport) of all displayed frames. 
We also measure the spatial and temporal quality variation, since these quality discontinuities can affect the perceptual quality. 
The {\it temporal quality discontinuity}  is the mean absolute difference between the rendering qualities of every two adjacent frames. 
To calculate the spatial variance of each frame, we measure the mean absolute difference between the rendering qualities of each tile and its neighboring tiles in the displayed FoV. Then the {\it spatial quality discontinuity} is the average of such spatial variance over every displayed frame.

\subsection{Evaluation results}
\label{sec:eval_result}
The PF region is set to cover an FoV size of $90^{\circ}\times90^{\circ}$. The streaming system adapts the sizes of FP+ and RI regions from the candidate sets. The candidate sizes of PF+ are $\{10^{\circ}, 20^{\circ}, 30^{\circ}, 40^{\circ}, 50^{\circ}\}$, while the candidate sizes of RI are $\{4, 8, 16, 32,64\}$ tiles.

%An RI size $= 0$ means that no RI regions are coded, which is preferred when the FoV prediction is very accurate. 

In Fig.~\ref{fig:trace}, the first plot is the entire bandwidth trace of 500-second duration we experimented on. The following plots are for a portion of the entire trace to show the details. 
From the bandwidth and the FoV traces, we can see the LSTM-based segment-level bandwidth prediction and the frame-level FoV prediction is very accurate, especially when the trace does not have random sudden changes. The accurate predictions lead to the high PF hit rate and the high frame delivery rate. 
From the traces of the region rates and sizes, we can see our proposed streaming system is able to adapt those parameters based on the FoV and bandwidth dynamics.  Specifically, we observe that the system tends to choose small RI and PF+ regions when the recent FoV predictions are  mostly accurate, while it tends to use larger RI and PF+ region sizes when the recent prediction accuracy drops. 
When the bandwidth suddenly drops to extremely low, we see the predicted bandwidth needs time to converge to the correct bandwidth. During the transient period, the frame delay increases, the frame delivery rate drops, and a larger RI size is used to increase the refresh frequency of the entire ERP. Note that better prediction algorithms can shorten this response time.

\begin{table*}
%\vspace{-4mm}
\small
\centering
\begin{tabular}{|l|c|c|c|c|c|c|c|c|c|c|}
\hline
 & \multicolumn{5}{c|}{{\it Trolley}} & \multicolumn{5}{c|}{{\it Chairlift}} \\ \hline
{\textbf{Metric}}&{\textbf{BM1}}&\textbf{{BM2}}&\textbf{{BM3}}&\textbf{{Prop.}}&\textbf{{Simp.}}&{\textbf{BM1}}&\textbf{{BM2}}&\textbf{{BM3}}&\textbf{{Prop.}}&\textbf{{Simp.}}\\ \hline
{\textbf{WS-PSNR in FoV (dB)}} & {36.17} & {38.23} & {44.25} & {48.41} & {48.35}      & {37.19} & {38.69} &{42.92} & {45.34} & {45.29}\\\hline
{Temporal discontinuity (dB)} & {0.236} & {0.204} & {0.298} & {0.229}  & {0.255}      & {0.177} & {0.146} &{0.277} & {0.159}  & {0.177}\\\hline
{Spatial discontinuity (dB)} & {0.274} & {0.002} & {0.005} & {0.203}  & {0.397}      & {0.192} & {0.001} &{0.006} & {0.199}  & {0.274}\\\hline
{\textbf{Average frame delay (ms)}} & {89.04} & {89.05} & {119.02} & {90.95}  & {90.72}      & {89.04} & {89.04} &{108.29} & {94.05}  & {93.73}\\\hline
{Delay STD/Average} & {0.272} & {0.272} & {0.467} & {0.308}  & {0.311}      & {0.270} & {0.270} &{0.392} & {0.312}  & {0.309}\\\hline
{Percentage of freeze frames (\%)} & {0.093} & {0.093} & {0.527} & {0.116}  & {0.113}      & {0.087} & {0.086} &{0.289} & {0.115}  & {0.112}\\\hline
{Average freeze duration (ms)} & {11.12} & {11.12} & {47.48} & {20.92}  & {20.74}      & {11.11} & {11.11} &{41.47} & {30.23}  & {24.49}\\\hline
{Display interval average (ms)} & {33.36} & {33.36} & {33.41} & {33.37} & {33.37}      & {33.36} & {33.36} &{33.39}  & {33.36} & {33.36}\\\hline
{Display interval STD (ms)} & {11.93} & {11.92} & {19.07} & {12.46} & {12.49}      & {11.93} & {11.92} &{19.07} & {12.91} & {12.86}\\\hline
{Average hit rate, PF (\%)} & {$N/A$} & {$N/A$} & {$N/A$} & {91.48} & {91.50}      & {$N/A$} & {$N/A$} &{$N/A$} & {91.03} & {91.08}\\\hline
{Average hit rate, PF+ (\%)} & {$N/A$} & {$N/A$} & {$N/A$} & {7.04} & {7.55}      & {$N/A$} & {$N/A$} &{$N/A$} & {6.97} & {7.62}\\\hline
{Average hit rate, RI (\%)} & {$N/A$} & {$N/A$} & {$N/A$} & {0.87} & {0.88}      & {$N/A$} & {$N/A$} &{$N/A$} & {0.97} & {0.95}\\\hline
{Average hit rate, total (\%)} & {54.33} & {99.86} & {99.90} & {99.39} & {99.93}      & {78.35} & {99.52} &{99.65} & {98.97} & {99.65}\\\hline
\end{tabular}
\caption{Streaming System Evaluation on {\it Trolley} (fixed camera) and {\it Chairlift} (moving camera).}
\label{tab:big_table}
\vspace{-4mm}
\end{table*}%

We compare the performance of the proposed and three benchmark systems using the metrics introduced in Sec.~\ref{sec:eval_matrics}. We report the average values over 48 users' traces for ``Trolley" and ``Chairlift" in Table~\ref{tab:big_table}. 
Compared with BM1 and BM2 using intro-coding only, we observe that the WS-PSNR of our proposed system is significantly higher (6-10dB higher), because the proposed system uses the region size/rate-adaptive inter-coding instead of using intra-coding only in the compared systems. However, those adapted rates of PF and PF+ regions bring a slightly higher spatial quality discontinuity (0.1-0.2dB higher) as a compromise. 
We observe BM2 leads BM1 in terms of the WS-PSNR because BM2's tile structure (same as the proposed) is finer and more flexible than BM1's vertical slice structure. When the PF center is close to the equator, this finer tile structure allows the system to generally encode and transmit a smaller area surrounding the predicted FoV. When the PF center is close to the north or south pole, systems using the tile structure are able to encode all tiles needed to cover the predicted FoV, while the fixed-width vertical slice of BM1 cannot cover the FoV horizontally spanned across the ERP, which also leads to a low FoV hit rate of BM1.
BM1, BM2, and the proposed system achieve similar good performance in terms of the average frame delay ($<100$ms with low variance), the probability of freeze ($<0.15\%$), and the duration of freeze ( $<1$ frame time), because they share the same proposed bandwidth prediction and bit budget allocation algorithms.
However, due to the variable bit rates of frames inside each segment brought by the varying coding time lapses, one small compromise of the proposed system is the slightly higher delay and freeze. Since the proposed system predicts the time lapse distribution in the new segment based on that in the previous segment, the prediction is not accurate for a segment when the FoV dynamics changes. This would increase the chance that the actual bit rate is higher than the allocated rate, leading to slightly increased frame delay and freezing probability.

Compared with BM3 using inter-coding, the proposed system achieves a significantly lower probability of freeze (60-78\% lower) and frame delay (14-28ms lower), because BM3 uses the traditional GOP coding structure which experiences periodic rate spikes when coding the I-frame in each segment.
The proposed system also has 2-4dB higher average WS-PSNR than BM3, because the proposed system optimizes the sizes and rates of PF and PF+, and codes fewer tiles using the intra-mode (determined by the adapted RI size and rate). 

To evaluate the benefit from adapting the region sizes, we also simulate a ``simplified system", which uses fixed PF+ size of $50^{\circ}$, and fixed RI size of 4 tiles, which is noted as ``Simp." in Table~\ref{tab:big_table}. We can see the performance degradation from the proposed to the simplified system is very small. Therefore, this simplified system might be more preferable for practical adoption, especially for low-power mobile devices. 

%because the scene content is more stationary  for "Trolley", it enjoys higher Q-R efficiency (see Fig.~\ref{fig:qr}), and consequently higher rendered quality.  
% ``Chairlift'' has lower quality primarily than "Trolley" because of its fast moving content and hence lower Q-R efficiency (see Figure ~\ref{fig:qr}).  ``Chairlift'' also suffers from slightly higher delay and freeze due to the more significant video rate variation caused by the more dynamic FoV traces. 

%We also have the system evaluation using the truncated-linear FoV prediction and RLS bandwidth prediction, which was described in \cite{mao2020low}. 

Our preliminary experiments reported in \cite{mao2020low} used truncated-linear FoV prediction and RLS bandwidth prediction. Compared to the results in \cite{mao2020low},  the current system using LSTM-based bandwidth prediction  reduces the average frame delay  by about 3-5ms and decrease the percentage of freeze frames by 60-70\%, and the total freeze duration by 10-15 ms. Meanwhile, the LSTM-based FoV prediction leads to increased 
%freeze frames by up to 60\% less, 
FoV hit rate (about 1\% higher), which in turn results in better rendered quality (about 0.1dB increase in WS-PSNR). Note that although the LSTM-based FoV prediction provides significant improvement in FoV hit rate in long-term prediction as shown in Fig.~\ref{fig:fov_lstm}, the proposed system enjoys short frame delay and typically only needs to predict the FoV of future 3-5 frames, for which the gain from the LSTM-based FoV prediction is limited.

%------------------------------------------------------------------------- 
\section{Conclusion}
\label{sec:conclusion}

In this paper, we developed a novel  temporal predictive coding scheme that can be integrated  into low-latency, FoV-adaptive streaming of interactive $360^\circ$ video. To stop error propagation due to frame losses as well as quality degradation in un-coded regions, we introduced 
rotating intra-regions without incurring bit rate spikes.  Accurate quality-rate models were derived to  explicitly consider the reduced coding efficiency resulting from the prolonged temporal prediction time lapse. Based on the quality rate models,  we investigated region size adaptation and rate allocation among regions at the segment  level to maximize the rendering quality. We further developed LSTM-based models for the prediction of future FoV and network throughput. Finally we designed a low-latency push-based streaming system that integrates the proposed video coding, FoV prediction, and throughput prediction schemes. Through extensive experiments, we demonstrated that the proposed system can reduce the mean end-to-end delay to below 100ms in the face of challenging network bandwidth variations. Compared to benchmark systems that always apply intra coding  on the predicted FoV regions, the proposed system provides substantial rendering quality improvement  (with  a gain of 7 to 11 dB in the average WS-PSNR), while maintaining similar  low latency and freeze probability. Relative to the benchmark system utilizing inter-coding with periodic intra-frames, the proposed system can significantly reduce the latency and the freeze probability, while also providing rendering quality improvement (up to 4dB).

%------------------------------------------------------------------------- acknowledgments
\begin{acks}
This material is based upon work supported by the National Science Foundation under Grant No. 1816500.
\end{acks}

\bibliographystyle{ACM-Reference-Format}
\bibliography{main}

\end{document}